\documentclass[12pt]{article}
\newcommand{\be}{\begin{equation}}
\newcommand{\ee}{\end{equation}}
\newcommand{\bea}{\begin{eqnarray}}
\newcommand{\eea}{\end{eqnarray}}

\usepackage [dvips]{graphicx}
\usepackage [footnotesize, hang, sc]{caption}
\usepackage[english]{babel}
\usepackage{amsfonts}
\usepackage{amsmath}
\usepackage[T1]{fontenc}
\usepackage[latin1]{inputenc}
\title{\bf The skewed multifractal random walk with applications to option smiles}
\author{Beno\^it Pochart\thanks{Centre de Math\'ematiques Appliqu\'ees,
Ecole Polytechnique, 91128 Palaiseau {\sc Cedex},
FRANCE (Email: pochart@cmapx.polytechnique.fr).} \, and
Jean-Philippe Bouchaud\thanks{Service de Physique de l'Etat Condens\'e,
Centre d'\'etudes de Saclay, Orme des Merisiers, 91191 Gif-sur-Yvette {\sc Cedex},
FRANCE (Email: bouchau@drecam.saclay.cea.fr).}
\thanks{Science \& Finance, Capital Fund Management,
109-111 rue Victor Hugo, 92532 Levallois {\sc Cedex},
FRANCE, http://www.science-finance.fr. } }
\date{}
\begin{document}
\maketitle
\abstract{We generalize the construction of the multifractal
random walk ({\sc mrw}) due to Bacry, Delour and Muzy to take into account the asymmetric
character of the financial returns. We show how one can include in this class
of models the observed correlation between past returns and future volatilities,
in such a way that the scale invariance properties of the {\sc mrw} are preserved.
We compute the leading behaviour of $q$-moments of the process, that behave as
power-laws of the time lag with an
exponent $\zeta_q=p-2p(p-1)\lambda^2$ for even $q=2p$, as in the symmetric {\sc mrw}, and
as $\zeta_q=p+1-2p^2\lambda^2-\alpha$ ($q=2p+1$), where $\lambda$ and $\alpha$ are parameters.
We show that this extended model reproduces the `{\sc harch}' effect or `causal cascade' reported
by some authors. We illustrate the usefulness of this `skewed' {\sc mrw} by computing the
resulting shape of the volatility smiles generated by such a process, that we compare to
approximate cumulant expansions formulas for the implied volatility. A large variety of smile
surfaces can be reproduced.}
\section{Introduction}
Volatility clustering in financial markets is a well known phenomenon.
More surprising is
the fact that volatility correlations are found to be long-ranged, 
and cannot be
characterized
by a single correlation time. Recent empirical works \cite{Lo,Ding,Stanley,Cont,BMD}
suggest that the volatility correlation
function of various assets (in particular stocks) actually decays as a power-law of the time lag, with a rather small exponent.
Furthermore, the volatility fluctuations are found to be close to (but not exactly) log-normal
\cite{Stanley2,Mantegna}.
These observations
have lead several authors \cite{Mandel,BMDshort,BMD}
to propose an interesting class of {\it multifractal models},
where the log-volatility is a Gaussian random variable with a correlation function that
decays in time as a logarithm. An explicit continuous time construction of such a process
was proposed and studied in \cite{BMD}, and was showed to exhibit strict multifractal
properties, in the sense that the even moments of the log-price difference scale with a
non trivial power of the time lag (more precise statements will be given below). This means
in particular that the kurtosis of the process decreases only very slowly with the time lag, in
contrast with most simple models of stochastic volatility, where the kurtosis drops
exponentially with time. This model is therefore of interest for option pricing, because
it is consistent with smiles that flatten only very slowly with time \cite{Book,PCB,Backus}.
The `multifractal random walk' ({\sc mrw}) constructed in \cite{BMD} is however
explicitly symmetric,
in the sense that all the odd moments of the log-price difference are strictly zero. The
same is true of the `cascade' construction of Mandelbrot, Fisher and Calvet
\cite{Mandel,Calvet1} (see also \cite{Calvet2} for an interesting causal
extension of this model).
On the other hand, an important stylized fact of the time series of stocks 
and stock indices is
the so called `leverage
effect': past price returns and future volatilities are negatively correlated \cite{leverage}.
This effect
was documented quantitatively in \cite{leverage,BMP,perello}. This effect is particularly strong for stock
indices, and induces a significant (negative) skewness in the distribution of price returns.
The aim of this paper is to study a generalization of the {\sc mrw} which accounts for
the leverage effect and the corresponding skewness. A particularly interesting class of models
preserves the multiscaling property of the {\sc mrw} and extends it to odd moments.
We discuss several financial applications of these models, in particular to option pricing.
In the presence of non zero skewness, the volatility smile itself becomes skewed. Approximate
theories, based on a cumulant expansion, are also discussed in this context.
\section{The Multifractal Random Walk ({\sc mrw})}
Let $(X_t, t \geq 0)$ be a stochastic process, with stationary increments.
We note by $\delta_\ell X_t$ the increments $X_{t+\ell}-X_t$ and by $M(q,\ell)$ the
$q-th$ moment  $E(|\delta_\ell X_t|^q)$. $(X_t, t \geq 0)$ is said to be a fractal process if
\begin{equation}\label{multiscaling}
M(q,\ell)=C_q \ell^{\zeta_q}.
\end{equation}
When the function $\zeta_q$ is linear in $q$, one speaks of monofractal process.
This is the case of the brownian motion for which $\zeta_q=q/2$,
and more generally of self-similar processes \cite{Taqqu-Samoro, intro_selfsimilar_processes},
for which the following equality in law holds
\begin{displaymath}
\delta_{b \ell}X_t = b^{H} \delta_\ell X_t,
\end{displaymath}
where $b$ is an arbitrary scaling factor.
When the function $\zeta_q$ is non linear in $q$, one speaks of multifractal process.
In this case, as emphasized in \cite{BMD}, Eq. (\ref{multiscaling}) can in fact only hold
in a certain `scaling regime', $\ell \ll T$.
The construction of \cite{BMD} is based on the following discretized process:
\begin{displaymath}
X_{\Delta t}(t)=\sum_{k=1}^{\frac{t}{\Delta t}} \delta X_{\Delta t}[k];
\qquad \delta X_{\Delta t}[k] = \epsilon_{\Delta t}[k]e^{\omega_{\Delta t}[k]}.
\end{displaymath}
For financial applications, $X_{\Delta t}(t)$ can be thought of as the logarithm of the
price at time $t$.
The quantities $\epsilon_{\Delta t}[k]$ and $\omega_{\Delta t}[k]$
are independent Gaussian variables with the following covariance structure:
\begin{eqnarray*}
E(\epsilon_{\Delta t}[j]\epsilon_{\Delta t}[k]) &=&\sigma ^2 \Delta t \delta_{j,k}\\
E(\omega_{\Delta t}[j]\omega_{\Delta t}[k]) &=&\lambda^2 \ln{\rho_{\Delta t}[|j-k|]}
\end{eqnarray*}
where
\begin{displaymath}
\rho_{\Delta t}[\ell]=\left \{ \begin{array}{ll} \frac{T}{(|\ell|+1)\Delta t} &
\textrm{when}\,\,\,|\ell|\leq \frac{T}{\Delta t}-1 \\
1 & \textrm{otherwise} \end{array} \right.
\end{displaymath}
$T$ is the integral time beyond which the multifractal scaling ceases to hold.
$\epsilon_{\Delta t}[k]$ is of mean $0$ and for the variance of the process to converge when $\Delta t \to 0$, one has to choose:
$$E(\omega_{\Delta t}[k])=-Var(\omega_{\Delta t}[k])=-\lambda^2 \ln{\frac{T}{\Delta t}}.$$
Under these conditions, it was shown in \cite{BMD}
that the even moments of the process are, in the limit $\Delta t \to 0$, given by:
\begin{displaymath}
M(2p,\ell) =K_{2p} \left(\frac{\ell}{T} \right)^{p-2p(p-1)\lambda^2}.
\end{displaymath}
with
\begin{displaymath}
K_{2p}=(2p-1)!! (\sigma^{2}T)^p \int_0^1du_1 \dots \int_0^1du_p \prod_{i<j}
|u_i-u_j|^{-4\lambda^2}.
\end{displaymath}
The explicit computation of the above integral can be found in \cite{BMD3}, that
shows that the moments are finite only for $q < q^*= 2+1/2\lambda^2$, beyond which these
moments are infinite for all $\ell$.
>From the definition of $\zeta_q$, we obtain:
\begin{displaymath}
\zeta_q=\frac{q}{2}(1-(q-2)\lambda^2)\,, \qquad q=2p.
\end{displaymath}
One can also compute the correlation of the local volatility, given by $\sigma[k] \equiv
\sigma e^{\omega_{\Delta t}[k]}$. It is easily shown that this correlation decays as
a power-law of the time lag \cite{BMD}:
\begin{displaymath}
E(\sigma[k]^q\sigma[k+\ell]^q) -E(\sigma[k]^q)^2 \propto \ell^{-q^2 \lambda^2}.
\end{displaymath}
Empirically, $\lambda^2$ is found to be rather small, in the range $0.02 -- 0.1$ \cite{BMD}.
\section{A Skewed Multifractal Random Walk ({\sc smrw})}
\subsection{Definition}
We now generalize the construction of \cite{BMD} in order to account for the leverage
effect, where the volatility is correlated with past price returns.
We first consider the following discretized model (we omit the $\Delta t$ in
$\epsilon$ and $\omega$ for simplicity):
\begin{displaymath}
\delta X_{\Delta t}[k]  = \epsilon[k] \ e^{\tilde{\omega}[k]} \qquad \tilde{\omega}[k]
\equiv \omega[k]-\sum_{i<k}K(i,k)\epsilon[i],
\end{displaymath}
where $K(i,k)$ is a certain kernel describing how the sign of the return at
time $i$ affects the (log)-volatility at a later time $k$.
We think of $K(i,k)$ as
being positive, and therefore the minus sign
accounts for the sign of the leverage
effect. In order to preserve the multiscaling properties of the
process, we choose the kernel $K(i,j)$ to decay as a power-law:
\begin{displaymath}
K(i,j) = \frac{K_0}{(j-i)^\alpha \Delta t^{\beta}}
\qquad (j>i).
\end{displaymath}
We now compute exactly the first moments of this process and
then give an approximate formula (to first order in $\Delta t$)
for the higher moments.
\subsection{The second moment}
Let $t=n\Delta t$, with $1 \ll n \ll N=\frac{T}{\Delta t}$. We find, using the notation
$\langle  ... \rangle \equiv E_{\epsilon,\omega}(...)$:
\begin{eqnarray*}
\langle  X_t^2 \rangle &=&\langle (\sum_{i=0}^{n-1}\delta
X_{\Delta t}[i])^2\rangle
=  \sum_{i=0}^{n-1} \langle \epsilon[i]^2\rangle \langle e^{2\tilde{\omega}[i]}\rangle \nonumber \\
&=& \sigma^2 \Delta t \sum_{i=0}^{n-1} \langle e^{2\omega[i]}\rangle \langle e^{-2\sum_{k<i}\frac{K_0}
{(i-k)^\alpha \Delta t^{\beta}}\epsilon[k]}\rangle \nonumber \\
&=& \sigma^2 \Delta t \sum_{i=0}^{n-1}  e^{2K_0^2\sigma^2 \Delta t^{1-2\beta}
\sum_{k<i}\frac{1}{(i-k)^{2\alpha}}}
\end{eqnarray*}
We now make the following assumptions:
\begin{eqnarray}
\alpha > \frac{1}{2} \nonumber \\
\sigma^2  K_0^2 \Delta t^{1-2\beta} \ll 1 \label{conditions1}
\end{eqnarray}
which, in particular, requires that $\beta \leq \frac{1}{2}$ when $\Delta t \to 0$.
The conditions (\ref{conditions1}) ensure that the sum
$\sum_{k<i}\frac{1}{(i-k)^{2\alpha}}=\sum_{\ell=1}^\infty \frac{1}{\ell^{2\alpha}}$
converges and that we can replace, to lowest order in $K_0$, the exponential terms in the
expression for $\langle X_t^2\rangle$, by
$1$. Under these assumptions, we simply have:
\begin{displaymath}
\langle X_t^2\rangle =\sigma^2 t,
\end{displaymath}
as for the simple Brownian motion, or for the symmetric {\sc mrw} considered in \cite{BMD}.
\subsection {The third moment}
The third moment, related to the skewness, explicitly reads:
\begin{eqnarray*}
\langle X_t^3\rangle  &=& \sum_{0\leq i,j,k<n}\langle \epsilon[i]\epsilon[j]\epsilon[k]
e^{\tilde{\omega}[i]}e^{\tilde{\omega}[j]}e^{\tilde{\omega}[k]}\rangle  \nonumber \\
&=& 3\sum_{0 \leq i<j<n}\langle \epsilon[i]\epsilon[j]^2e^{\tilde{\omega}[i]+2\tilde{\omega}[j]}\rangle \nonumber \\
&=& 3\sigma^2 \Delta t\sum_{0 \leq i<j<n}\langle e^{\omega[i]+2\omega[j]}\rangle
\langle \epsilon[i]e^{2K(i,j)\epsilon[i]}\rangle  \nonumber \\
&& \langle e^{\sum_{k<i} \left (K(k,i)+2K(k,j) \right )\epsilon[k]}\rangle
\langle e^{2\sum_{i<k<j}K(k,j)\epsilon[k]}\rangle  \\ \nonumber
\end{eqnarray*}
Within the conditions (\ref{conditions1}),
the exponential terms (i.e. $\langle e^{K(k,j)\epsilon_k}\rangle $) can again be set to
1 to first order,
provided the sums converge, which is the case whenever  $\alpha > 1/2$.
Using the equality $\langle \epsilon e^{\lambda \epsilon}\rangle =
\lambda \sigma^2 e ^{\frac{\lambda^2 \sigma^2}{2}}$, we find
\begin{eqnarray}
\langle X_t^3\rangle  &\simeq &  6(\sigma^2 \Delta t)^2\sum_{0 \leq i<j<n}
K(i,j)\rho_0^{-\frac{\lambda^2}{2}}\rho_{i,j}^{2\lambda^2}\nonumber \\
& = & -6K_0\sigma^4 T^{\frac{3\lambda^2}{2}}
\Delta t^{\alpha-\beta +\frac{\lambda^2}{2}}\sum_{0 \leq i<j<n}
\frac{\Delta t}{((j-i)\Delta t)^\alpha}
\frac{\Delta t}{((j-i+1)\Delta t)^{2\lambda^2}}\nonumber \\
&\stackrel{\Delta t \to 0}\simeq &
-6K_0\sigma^4 T^{\frac{3\lambda^2}{2}}
\Delta t^{\alpha-\beta +\frac{\lambda^2}{2}}
\iint \limits_{0<u<v<t}\frac{du dv}{|u-v|^{\alpha+2\lambda^2}} \label{calcul3}
\end{eqnarray}
One easily proves that, under the condition $\nu_2=\alpha +2\lambda^2<1$,
\begin{displaymath}
\iint \limits_{0<u<v<t}\frac{du dv}{|u-v|^{\alpha+2\lambda^2}}=
\frac{t^{2-\nu_2}}{(2-\nu_2)(1-\nu_2)},
\end{displaymath}
so that our final result is:
\begin{equation}\label{3rdmoment}
\langle X_t^3\rangle  \approx - \frac{6K_0\sigma^4 T^{\frac{3\lambda^2}{2}}}{(2-\nu_2)(1-\nu_2)}
\Delta t^{\mu}t^{2-\nu_2},
\end{equation}
where $\mu = \alpha-\beta +\frac{\lambda^2}{2}$.
Therefore, the third moment grows as a power of the time lag $t$, with an exponent
$\zeta_3=2-\nu_2$.
Due to the conditions (\ref{conditions1}), $\mu$ is always positive
and therefore $\langle X_t^3\rangle $ vanishes in the limit $\Delta t \to 0$. In other words,
the skewness of the continuous-limit process disappears.
In practice, however, the elementary time scale $\Delta t$, beyond which price
increments are uncorrelated, is equal to several seconds, even in futures markets, and
to several minutes in stocks markets.
>From a theoretical perspective, however, it would be interesting to
construct a multifractal process which remains skewed in the continuous time limit.
We will come back to this point in section \ref{SEMRW}.
\subsection{Higher order moments}
As $q$ grows, the algebra of the explicit computation of the $q-th$ moment rapidly becomes messy,
and the exact result cannot be computed. However, as we show in appendix, one can still
compute the dominating term in the limit $\Delta t \to 0$.
\subsubsection{Odd moments}
We want to evaluate the generic moment:
\begin{equation}
\langle X_t^{2p+1}\rangle  = \sum_{0 \leq i_1,\dots, i_{2p+1}<n}
\langle \epsilon[i_1] \dots \epsilon[i_{2p+1}] \, e^{\tilde{\omega}[i_1]}
\dots e^{\tilde{\omega}[i_{2p+1}]}\rangle .
\label{moment_impair}
\end{equation}
The sum over $2p+1$ variables contains many different terms but, as we justify in the
Appendix, the dominating one is:
\begin{equation}
\mathcal{D}_{2p+1}=\sum_{0 \leq j_1,\dots, j_{p+1}<n}\langle \epsilon[j_1] \epsilon[j_2]^2
\dots \epsilon[j_{p+1}]^2 e^{\tilde{\omega}[j_1]+2\tilde{\omega}[j_2]+ \dots +
2\tilde{\omega}[j_{p+1}]}\rangle  \label{dominant_impair}
\end{equation}
where all the variables are paired two by two but one.
For small $K_0$, we finally find:
\begin{equation}
\langle X_t^{2p+1}\rangle  \simeq  -\mathcal{M}_{2p+1} K_0 \Delta t^\mu \, t^{p+1-2p^2\lambda^2-\alpha}, \label{oms}
\end{equation}
where $\mathcal{M}_{2p+1}$ is a positive numerical  prefactor, and $\mu$ was defined
above. Therefore, we find that all odd moments scale as a power of the time lag $t$, with
$\zeta_q=p+1-2p^2\lambda^2-\alpha$ ($q=2p+1$). In this sense, the model considered
is multifractal for odd moments as well.
\subsubsection{Even moments}
We now want to evaluate the generic even moment
\begin{displaymath}
\langle X_t^{2p}\rangle  = \sum_{0 \leq i_1,\dots, i_{2p}<n}
\langle \epsilon[i_1] \dots \epsilon[i_{2p}] \, e^{\tilde{\omega}[i_1]+ \dots +
\tilde{\omega}[i_{2p}]}\rangle
\end{displaymath}
Using similar arguments, detailed in the appendix, we can show that the
dominating term is simply:
\begin{displaymath}
\mathcal{D}_{2p}= \sum_{0 \leq j_1,\dots,j_{p}<n}\langle \epsilon[j_1]^2
\dots \epsilon[j_{p}]^2\, e^{2\tilde{\omega}[j_1]+ \dots  + 2\tilde{\omega}[j_{p}]}\rangle
\end{displaymath}
and that we retrieve, for $K_0$ small, the result of \cite{BMD}:
\begin{displaymath}
\langle X_t^{2p}\rangle  \simeq  \mathcal{M}_{2p}\, t^{p-2p(p-1)\lambda^2}.
\end{displaymath}
Hence, we find that a power-law leverage correlation function, introduced as we have
proposed, does not affect even moments (to lowest order in $K_0$), but give to the
odd moments a multifractal behaviour.
\section{An extension to slowly decaying kernels} \label{SEMRW}
As we have noticed above, a condition for the convergence of sums appearing
in the exponentials is $\alpha > 1/2$. A way to relax this is to add a large time cut-off
to the power-law kernel $K(i,j)$. More precisely, we set:
\begin{displaymath}
K(i,j) = \frac{K_0 \, e^{-\Gamma(j-i)}}{(j-i)^\alpha \Delta t^{\beta}} \qquad (j>i),
\end{displaymath}
where $\Gamma \sim \Delta t/T \ll 1$.
We now consider the case condition $\alpha < 1/2$, but impose:
\begin{equation}
\Gamma^{2\alpha -1} K_0^2\Delta t^{1-2\beta} \ll 1.\label{conditions2}
\end{equation}
This ensures that the sum in the exponential terms is convergent and remains small.
Following the same route as above, the third moment reads
\begin{displaymath}
\langle {X}_t^3\rangle
\stackrel{\Delta t \to 0}\simeq -6K_0\sigma^4 T^{\frac{3\lambda^2}{2}} \Delta t^{\mu}
\iint \limits_{0<u<v<t}\frac{e^{-\frac{|u-v|}{T}}du dv}{|u-v|^{\alpha+2\lambda^2}}.
\end{displaymath}
Note that the double integral over $u,v$ converges provided $\alpha+2\lambda^2 < 1$.
Under the condition $t \ll T$, the exponential term can be dropped and we recover the
same result as above, Eq.~(\ref{3rdmoment}), with $\mu = \alpha-\beta +\frac{\lambda^2}{2}$.
Unfortunately, (\ref{conditions2}) with $\Gamma \sim \Delta t/T$ impose that $\alpha-\beta \geq 0$
in the limit $\Delta t \to 0$. Therefore, the exponent $\mu$ is strictly positive and the
skewness again disappears in the continuous time limit.
On the other hand, the possibility of choosing $\alpha < 1/2$ leads to the interesting possibility
of a skewness that {\it grows} with time. Indeed, the third moment (given by  Eq.~(\ref{3rdmoment})),
divided by the third power of the volatility, scales as $t^{1/2-\nu_2}$, with $\nu_2=\alpha +2\lambda^2$.
Small values of $\alpha$ therefore leads to a growing skewness (at least for $t \ll T$), whereas $\alpha > 1/2$ necessarily leads to a skewness that decays with time.
\section{Numerical results}
\subsection{Methodology}
\subsubsection{Simulation}
In order to generate correlated Gaussian random variables (the $\omega$'s),
we follow a quite standard procedure explained in \cite{Beran}, based on
the utilization of the Fast Fourier Transform \cite{Numerical}.
\\The major difficulty is to efficiently compute the skewness convolution
$\sum_{i<k}K(i,k)\epsilon[i]$. We first generate $2n$ independent Gaussian variables $\eta[i]$,
and periodize our series as:
\begin{displaymath}
\epsilon[i]=\left \{ \begin{array}{ll} \eta[i] & \textrm{if } 0 \leq i <k \\
\eta[i+2n] & \textrm{if } k-2n\leq i <0 \end{array} \right .
\end{displaymath}
and we define
\begin{displaymath}
\bar{K}(i, k)=\left \{ \begin{array}{ll} K(i,k) & \textrm{if } 0 < k-i \leq n \\
0 & \textrm{otherwise} \end{array} \right .
\end{displaymath}
We can now evaluate the skewness convolution by $\sum_{i=k-2n}^{k-1}\bar{K}(i, k)\epsilon[i]$,
again using Fast Fourier Transforms. The use of $n$ `pseudo' past variables and the cut-off
imposed to $\bar{K}(i, k)$ are essential to avoid spurious correlations between past volatilities
and future returns. The systematic use of the Fast Fourier Transform makes the whole
procedure quite efficient and allows us to simulate long series of our process
(in the applications we used series of $2^{13}$ variables but longer series can be obtained).
\subsubsection{Evaluation of the moments}
\begin{figure}[!htbp]
  \begin{center}
    \includegraphics[scale=0.35]{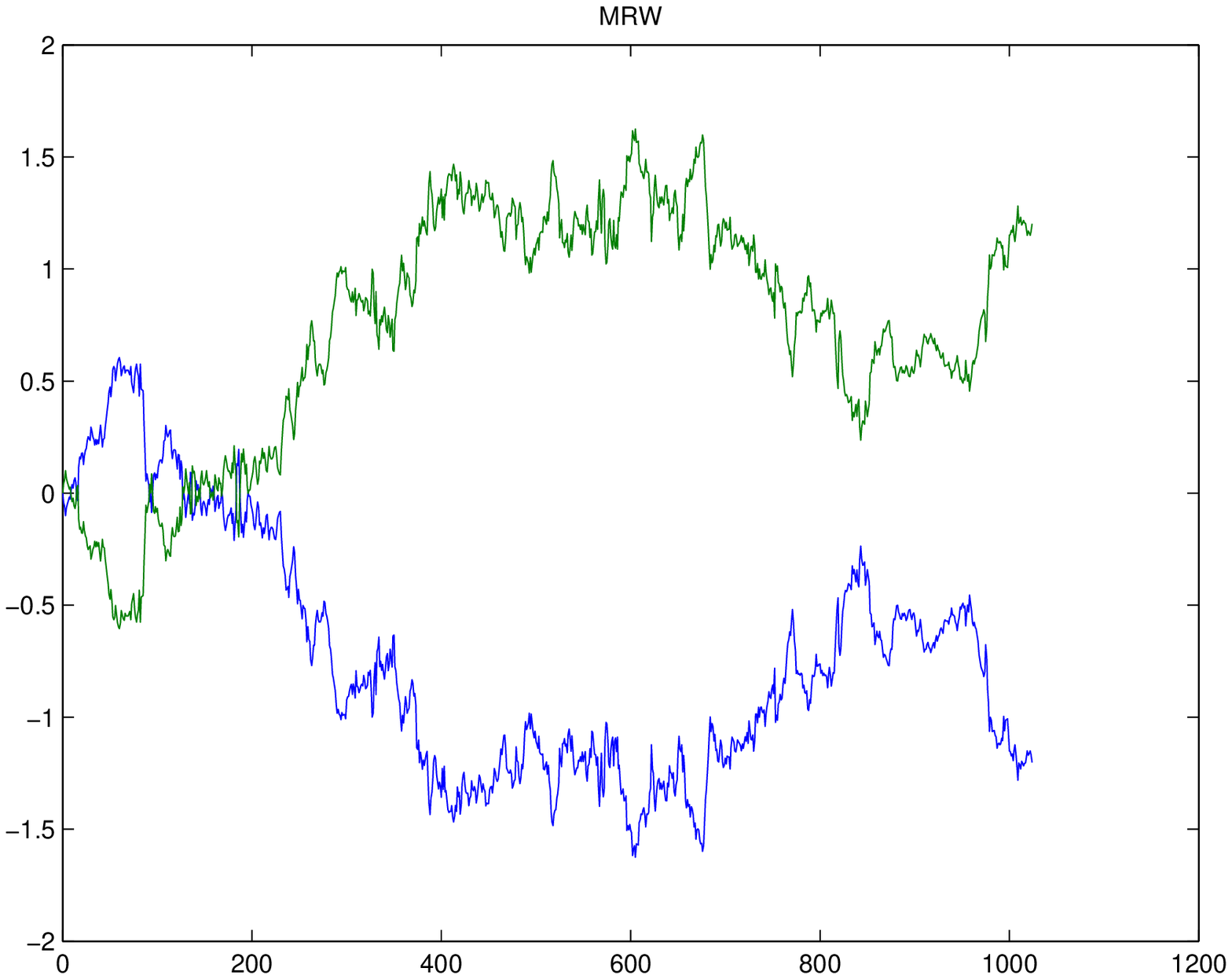} \includegraphics[scale=0.35]{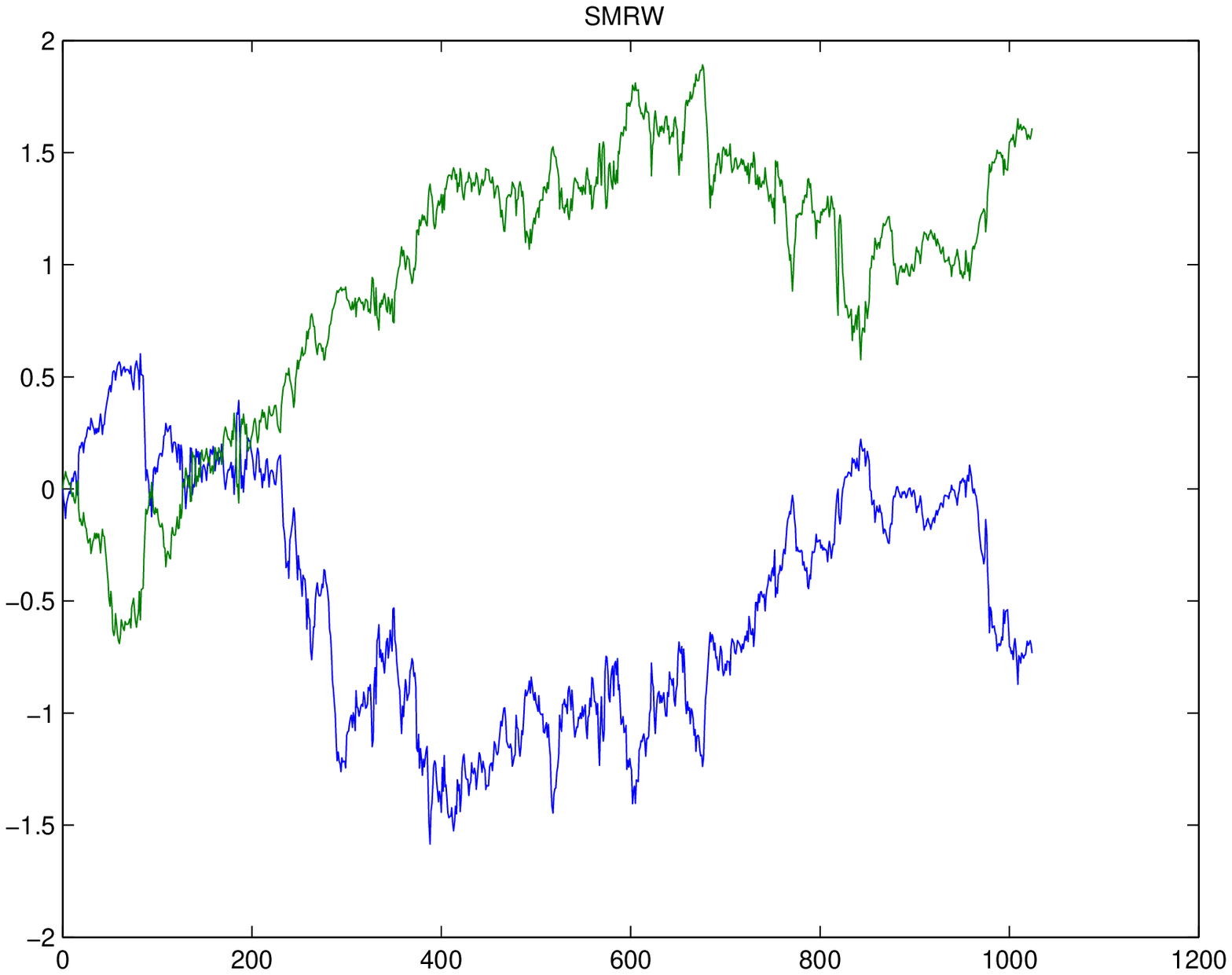}
    \caption{\label{sym} A realization of a {\sc mrw} and its ``mirror-image'' (left), the same for
    {\sc smrw} (right). In this case we observe the influence of the kernel $K$ which induce asymmetry in the process.}
  \end{center}
\end{figure}

We have evaluated the moments by averaging over a large number
of realizations of the process. This is not such an easy task, because the quantities
we deal with can be small and the convergence can be very slow due to the
existence of long-range correlations in the variables $\{\omega[i]\}$, and the
slow power-law decay of the kernel $K(i,j)$. In order to eliminate spurious skewness
effects and to be more accurate, we have systematically averaged over the processes
generated by the variables $\{\epsilon[i],\,\omega[i]\}$ and by the variables $\{-\epsilon[i],\,\omega[i]\}$. In the absence of asymmetry
(i.e. when $K(i,j)\equiv 0$), the second process is exactly the ``mirror image'' of the first one and therefore all odd moments are strictly zero. Conversely in the presence of asymmetry, the second process is no longer the ``mirror image'' of the first one (see Fig \ref{sym}), and one finds non zero odd moments. By proceeding this way, we are sure that any non zero
result can indeed be attributed to the leverage kernel.
\subsection {Some results}
\begin{figure}[!h]
  \begin{center}
    \includegraphics[scale=0.5]{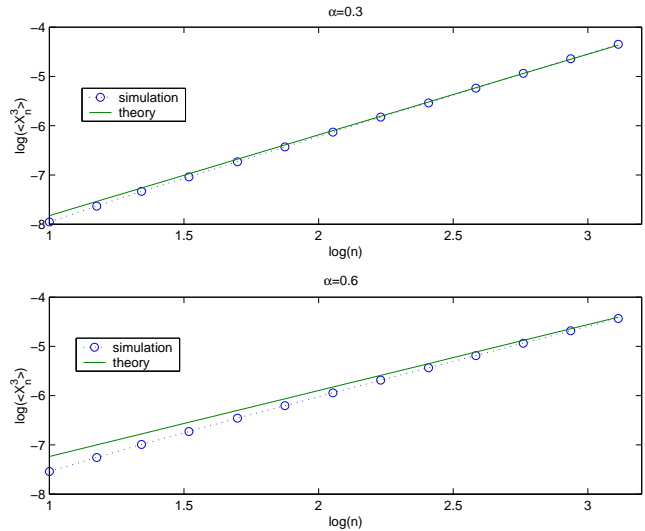}
    \caption{ \label{m3} Logarithm (base 10) of the $3^{rd}$ moment of the processes, for
$\alpha=0.3$ (top) and $\alpha=0.6$ (bottom), and $K_0=0.1$. The
adequation between simulations and theory is extremely good for $\alpha=0.3$ (top),
but less convincing for $\alpha=0.6$ (bottom). This discrepancy can be simply explained
by the faster convergence of the Riemann sums in the case
$\alpha=0.3$ than in the case $\alpha=0.6$ (see Fig. \ref{differences}}
  \end{center}
\end{figure}

\begin{figure}[!h]
  \begin{center}
    \includegraphics[scale=0.35]{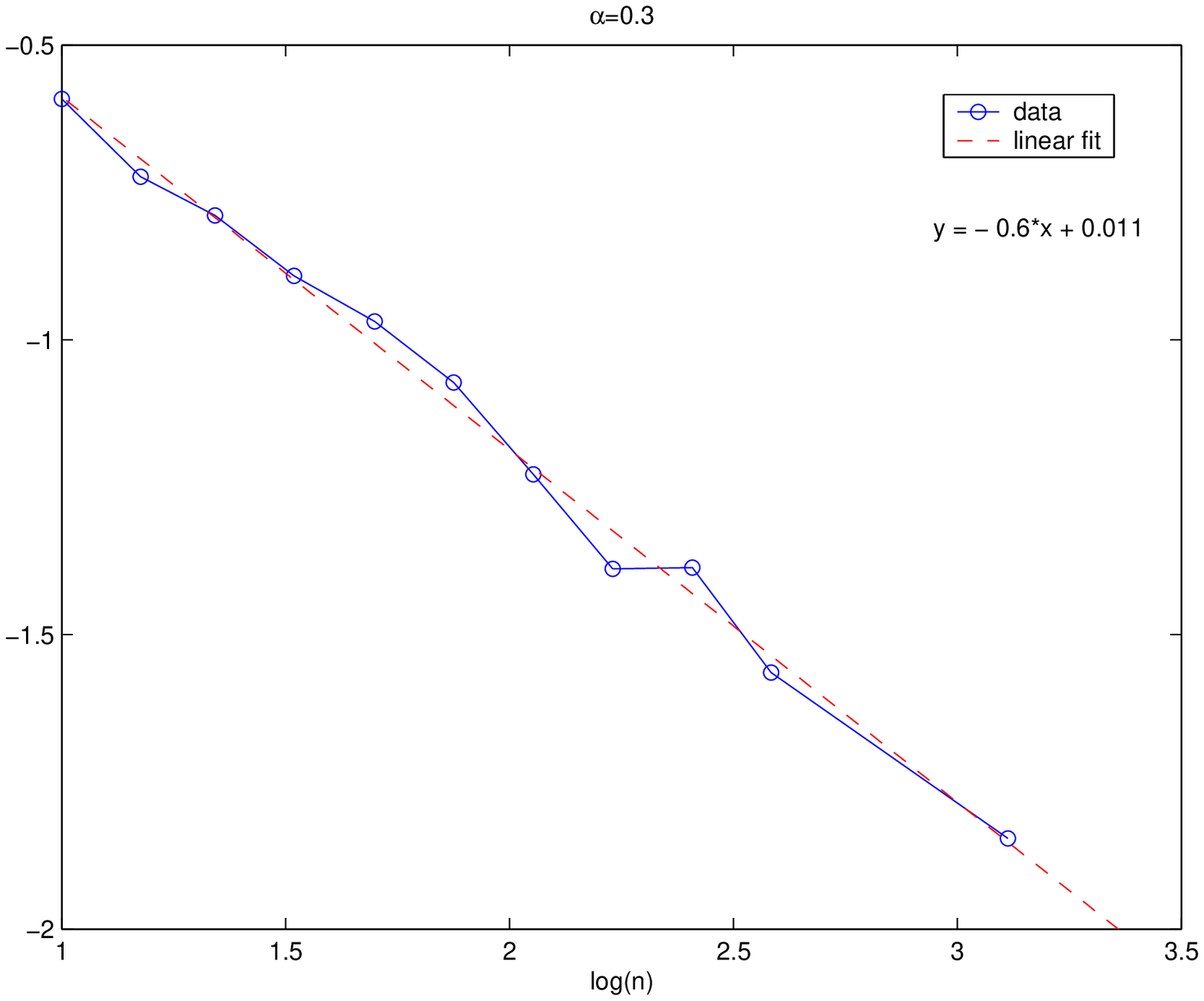} \includegraphics[scale=0.35]{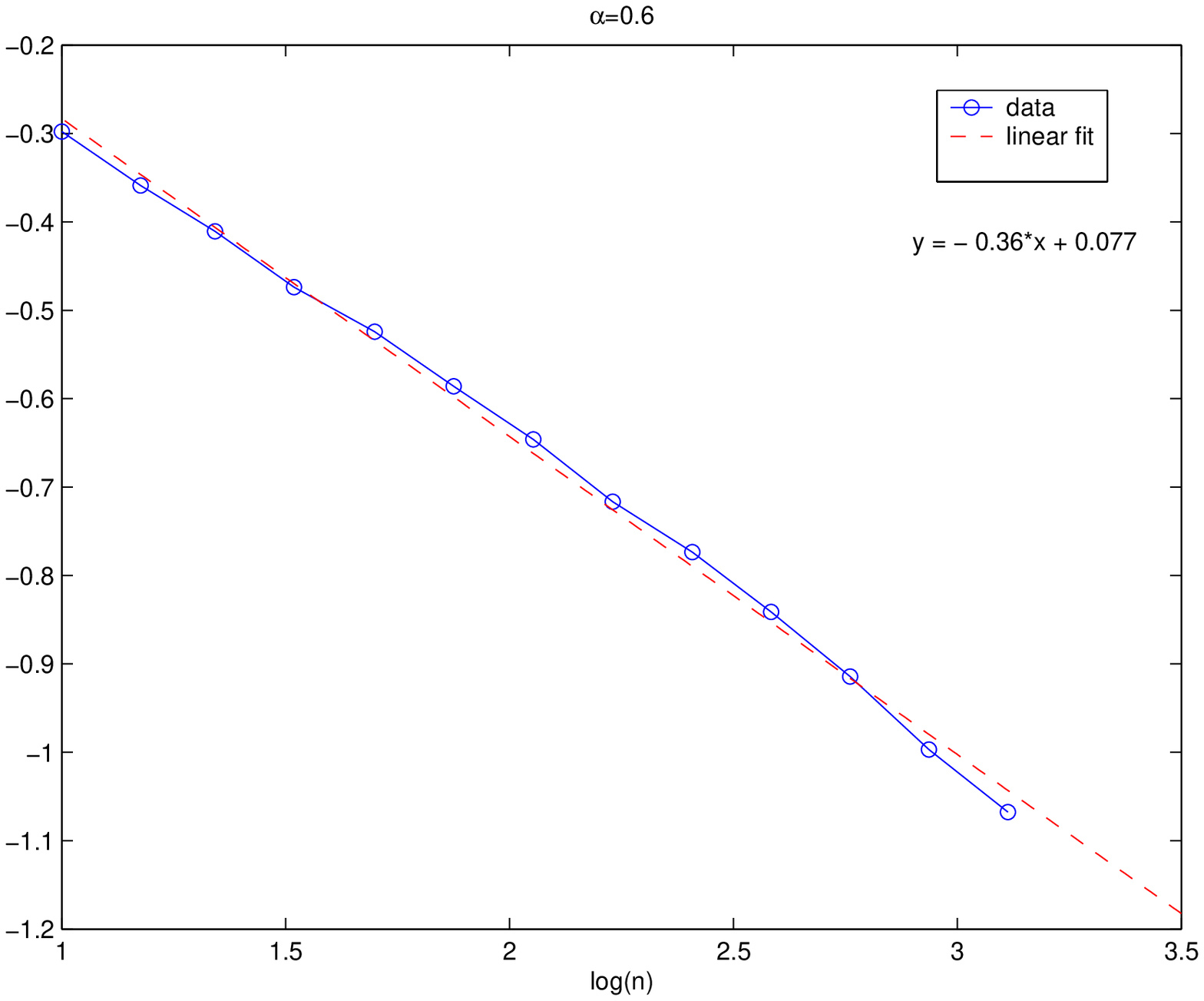}
    \caption{\label{differences} Logarithm (base 10) of the rescaled
    difference between the measured $3^{rd}$ moment of the process
    and the theoretical one (${\langle X^3_{th} \rangle- \langle X^3_{meas}\rangle}/
    {\langle X^3_{th}\rangle}$) as a
    function of the logarithm of the time lag ($n=t/{\Delta t}$). This graph
    confirms that the rescaled difference behaves like a power-law of the time lag,
    with an exponent close to $1-\nu_2$, that depends on $\alpha$.
    These corrections become large when $\nu_2 \to 1$ (for example for $\alpha=0.6$)
    and explain the discrepancy observed in Fig. \ref{m3}}
  \end{center}
\end{figure}

The results we present are based on 10 000 independent realizations of the process.
A total of $2^{13}$ variables were used for each series. The parameters we used are $T=4000 \Delta t$, $\lambda=0.175$, corresponding to $\lambda^2 \simeq 0.03$. We chose two different values of the
kernel exponent: $\alpha=0.3$ with $\beta=\alpha+\lambda^2/2$ and $\alpha=0.6$ with $\beta=0.45$.
The strength of the leverage effect is taken to be $K_0=0.1$, small enough to trust our
approximate formulas.
We have evaluated the different moments for times ranging from $t=10 \Delta t$ to $t =1300\Delta t$,
i.e. in a region where $\Delta t \ll t \ll T$. We have plotted the logarithm of the different moments with as a function of the logarithm of the time
lag. We expect to obtain a straight line with a slope given by $\zeta_q$.
 \begin{figure}[!h]
  \begin{center}
    \includegraphics[scale=0.35]{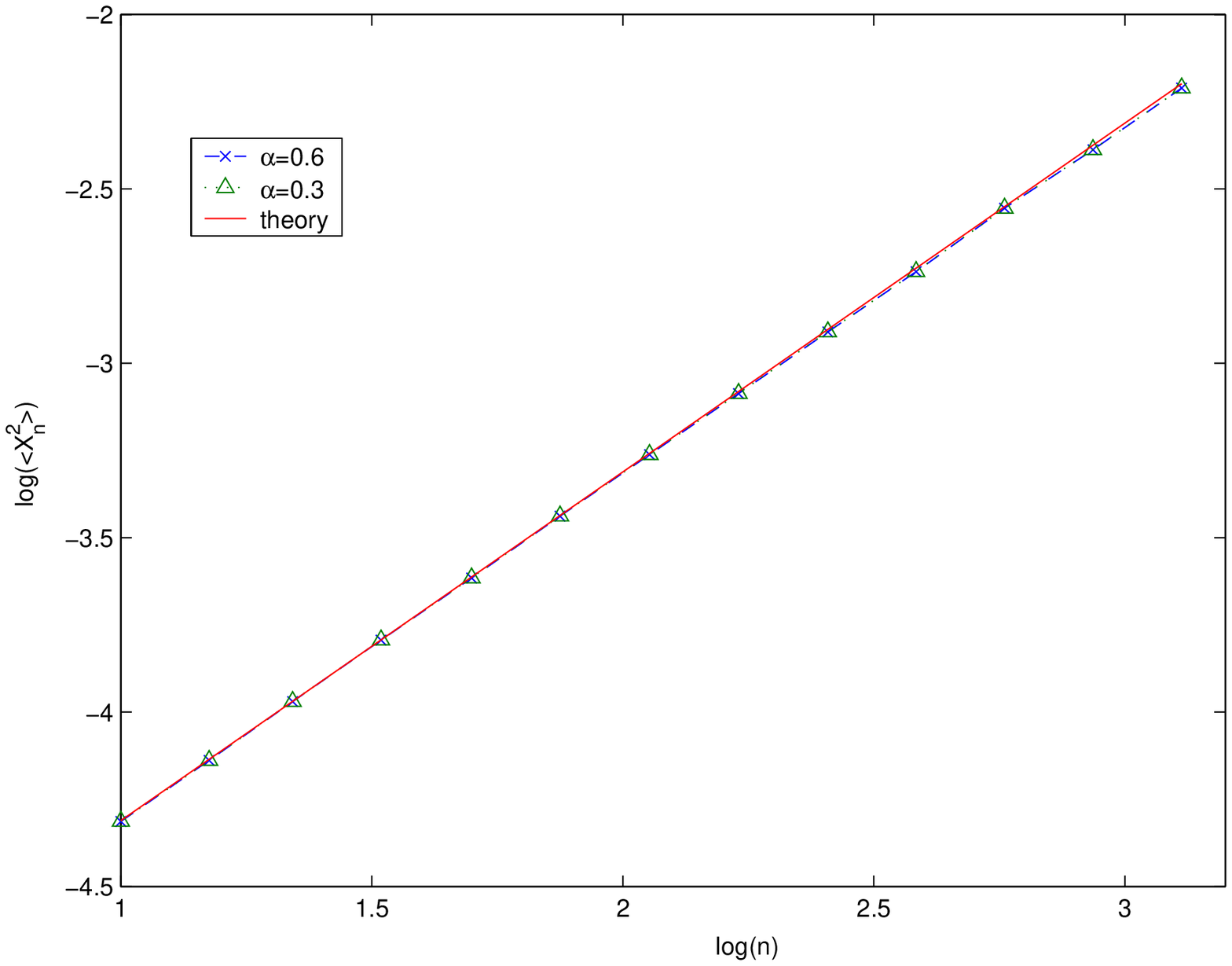} \includegraphics[scale=0.35]{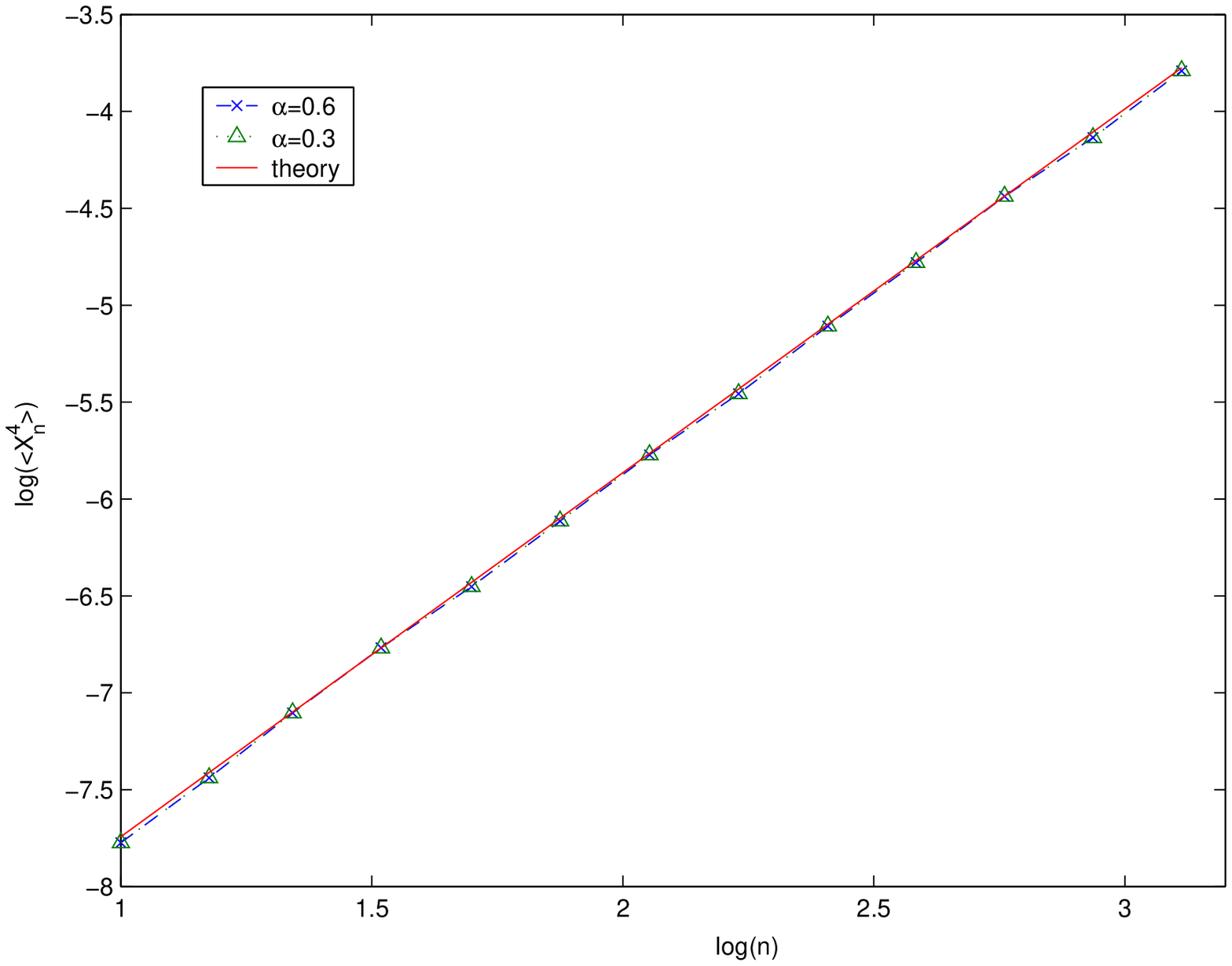}
    \caption{ \label{m24} Logarithm (base 10) of the $2^{nd}$ (left) and $4^{th}$ (right)
moments of the process, with $\alpha=0.6$ ($\times$) and $\alpha=0.3$ ($\triangle$)), as
a function of the logarithm of the time lag. We have shown for comparison the theoretical prediction, which is independent of $\alpha$. The agreement is extremely good over the whole time region, both
for the exponent $\zeta_q$ and the prefactor.
    }
  \end{center}
\end{figure}
We indeed observe a nice power-law scaling for the second and fourth moment of the process (Fig. \ref{m24}), with a perfect adequation between the theoretical prediction and the simulations.
The agreement is again very good for the third moment (Fig. \ref{m3}), in particular
for $\alpha=0.3$. There is a systematic difference for $\alpha=0.6$, that tends to disappear
for large time
lags. This is due to the fact that corrections to our asymptotic formulas tend to be stronger
when $\alpha$ increases. Indeed, in the above calculations (\ref{calcul3}), we have set:
$$
\sum_{0 \leq i<j<n}
\frac{\Delta t}{((j-i)\Delta t)^\alpha}
\frac{\Delta t}{((j-i+1)\Delta t)^{2\lambda^2}} \approx \iint \limits_{0<u<v<t}\frac{du dv}{|u-v|^{\alpha+2\lambda^2}}.
$$
Although this is correct in the limit $\Delta t/t \to 0$, there are corrections of the order
of $(\Delta t/t)^{1-\nu_2}$, which persist when $\nu_2 \to 1$. Let us note that this finite
time correction is very small for the even moments, since in this case $\nu_2$ is
replaced by $4\lambda^2$, which is very small (see Fig. \ref{differences}).
\section {Comparison with empirical data}
The comparison between the predictions of the (symmetric) {\sc
mrw} and real financial data was discussed in details in
\cite{BMD}. In particular, the way to calibrate the parameters
$\lambda^2$ and $T$ is to study the correlation function of the
log-volatility, that should behave as the log of the time lag up
to time $T$, with a slope which is precisely $\lambda^2$. While
$\lambda^2$ is quite well determined (its typical value is 0.03),
the `integral time' $T$ is much less precisely calibrated (it is found
to be $\approx 2$ years). The elementary time interval $\Delta t$ is
given by the chosen discretization of the time series, provided it
is larger than the correlation time of the returns (otherwise the
assumption that $\langle \epsilon[i] \epsilon[j] \rangle =
\delta_{i,j}$ is not justified). For example, for daily data, $\Delta t=1$ day. 

As far as skewness
is concerned, the following `leverage' correlation
function was proposed and studied in \cite{BMP}:
\begin{displaymath}
{\mathcal{L}}(i,j)=\frac{\langle \delta X[i]\delta X[j]^2 \rangle}{\left(
\langle \delta X[k]^2 \rangle \right)^2} \qquad i<j\,.
\end{displaymath}
Using the above definition we find, in the limit $K_0^2 \sigma^2 \Delta t \ll 1$,
the following relation:
\begin{equation}\label{eqleverage}
{\mathcal{L}}(i,j)
\equiv -2\left(\frac{T}{\Delta t}\right)^{\frac{3\lambda^2}{2}}
\frac{K(i,j)}{|i-j|^{2\lambda^2}},
\end{equation}
that allows in principle to calibrate $K(i,j)$ using empirical
data, once $\lambda^2$ and $T$ are fixed. (Note that only the
combination $K_0 T^{\frac{3\lambda^2}{2}}$ appears in the above
formulae, so that the uncertainty on $T$ does not transpire on the
skewness). The leverage correlation function is however found to
be, in the case of stock indices, close to a pure exponential in
$|i-j|$, with a decay time of ten days or so \cite{BMP}, in
contrast with Eq.~(\ref{eqleverage}). For individual stocks,
however, the decay time is much longer. The introduction of well
defined time scale would ruin the scaling properties of the {\sc
mrw} process. We choose to preserve the scaling properties of the
process for three reasons: (a) a strict multifractal skewed
process is an interesting process in its own right, that may have
applications outside finance (for example in turbulence \cite{turbulence}), (b) the scaling properties of the
{\sc smrw} might enable one to find exact analytic formulas for,
e.g. probability distributions, and (c) the slow decay of the
leverage function allows one to obtain long time persistent skews,
and is useful in the context of option pricing (see section
\ref{smiles} below). In other words, even if the historical time
series do not exhibit a long ranged leverage correlation function,
the implied distributions, consistent with option smiles, might do
so.
\section{Volatility asymmetry}
In \cite{AMS},
an interesting time asymmetry in financial series was detected using a wavelet analysis
(see also \cite{HARCH} for related work).
According to these authors, ``the volatility at large scales influences causally (in the future)
the volatility at shorter scales'', whereas the converse effect is much weaker.
This finding is actually deeply related to the leverage effect \cite{Muzy-private}
and our model, with its extended time correlations between returns and volatility, should be able to reproduce it, as we show now.
One can quantify the effect reported in \cite{HARCH,AMS}
by computing a correlation between volatilities
at different scales $n$ and $m$, defined as:
\begin{eqnarray*}
C(n,m)&=&\langle  \left(\sum_{i=0}^{n-1}\delta X[i]\right)^2
\left(\sum_{j=n}^{n+m}\delta X[j]\right)^2\rangle
\\
&=&\sum_{i=0}^{n-1}\sum_{j=n}^{n+m}
\langle \delta X[i]^2\delta X[j]^2\rangle+
2\sum_{0\leq i_1<i_2<n}\sum_{j=n}^{n+m}\langle \delta X[i_1]\delta X[i_2]\delta X[j]^2\rangle.
\end{eqnarray*}
Although this quantity is not exactly the one considered in \cite{AMS},
one expects that it behaves very similarly.
The first term $C_1$ is easy to compute and is equal, in the limit $\Delta t \to 0$, to:
\begin{eqnarray*}
C_1&=&\sigma^4 T^{4\lambda^2}
\int_{u=0}^\tau\int_{v=\tau}^{\tau+\tau'}\frac{du\,dv}{|u-v|^{4\lambda^2}}
\\
&=&\frac{\sigma^4 T^{4\lambda^2}}{(1-4\lambda^2)(2-4\lambda^2)}[(\tau+\tau')^{2-4\lambda^2}
-\tau^{2-4\lambda^2}-{\tau'}^{2-4\lambda^2}],
\end{eqnarray*}
where $\tau=n \Delta t$ and $\tau'=m\Delta t$. This result is symmetric in $\tau$ and $\tau'$,
and therefore cannot explain the asymmetry. The second term, on the other hand, is equal to:
\begin{eqnarray}
C_2&=&2\sigma^6 K_0^2 T^{4\lambda^2} \Delta t ^{\lambda^2+2\alpha -2\beta}\int_{0\leq u_1<u_2<\tau \leq v
<\tau+\tau'}\frac{2}{(v-u_2)^\alpha}\\ \nonumber
& &
\bigg[\frac{1}{(u_2-u_1)^\alpha}+
\frac{2}{(v-u_1)^\alpha}\bigg]
\frac{du_1\,du_2\,dv}{(u_2-u_1)^{\lambda^2}(v-u_1)^{2\lambda^2}(v-u_2)^{2\lambda^2}}.
\label{correlation_vol_passe-vol_futur}
\end{eqnarray}
We now change variables and set:
\begin{eqnarray*}
u_1 &=& x_1 \tau \\
u_2 &=& x_2 \tau \\
v   &=& \tau+y\tau',
\end{eqnarray*}
and assume first that $\tau' =\epsilon \tau$ with $\epsilon \ll 1$, i.e. we compare
future short scale volatilities with past large scale volatilities. One finds:
\begin{displaymath}
C_2 \propto  K_0^2 \epsilon \,\tau^{3-2\alpha-5\lambda^2}.
\end{displaymath}
Now, the opposite case where $\tau =\epsilon \tau'$ with $\epsilon \ll 1$,
i.e. where we compare
future large scale volatilities with past small scale volatilities leads to:
\begin{displaymath}
C_2 \propto  K_0^2 \epsilon^{2-\alpha-\lambda^2} \, \tau^{3-2\alpha-5\lambda^2}.
\end{displaymath}
The ratio of the second indicator to the first is therefore
equal to $\epsilon^{1-\alpha-\lambda^2}$,
which goes to zero provided that $\alpha+\lambda^2 < 1$. In this case, the asymmetry indeed
has the sign found in \cite{AMS},
that is, the correlation between past volatility
at short scale and future volatility at large scale is
indeed weaker than the correlation between past volatility at
large scale and future volatility at short scale.
\section{An application to option pricing: skewed volatility smile}\label{smiles}
\begin{figure}[!htbp]
  \begin{center}
    \includegraphics[angle=270,scale=0.26]{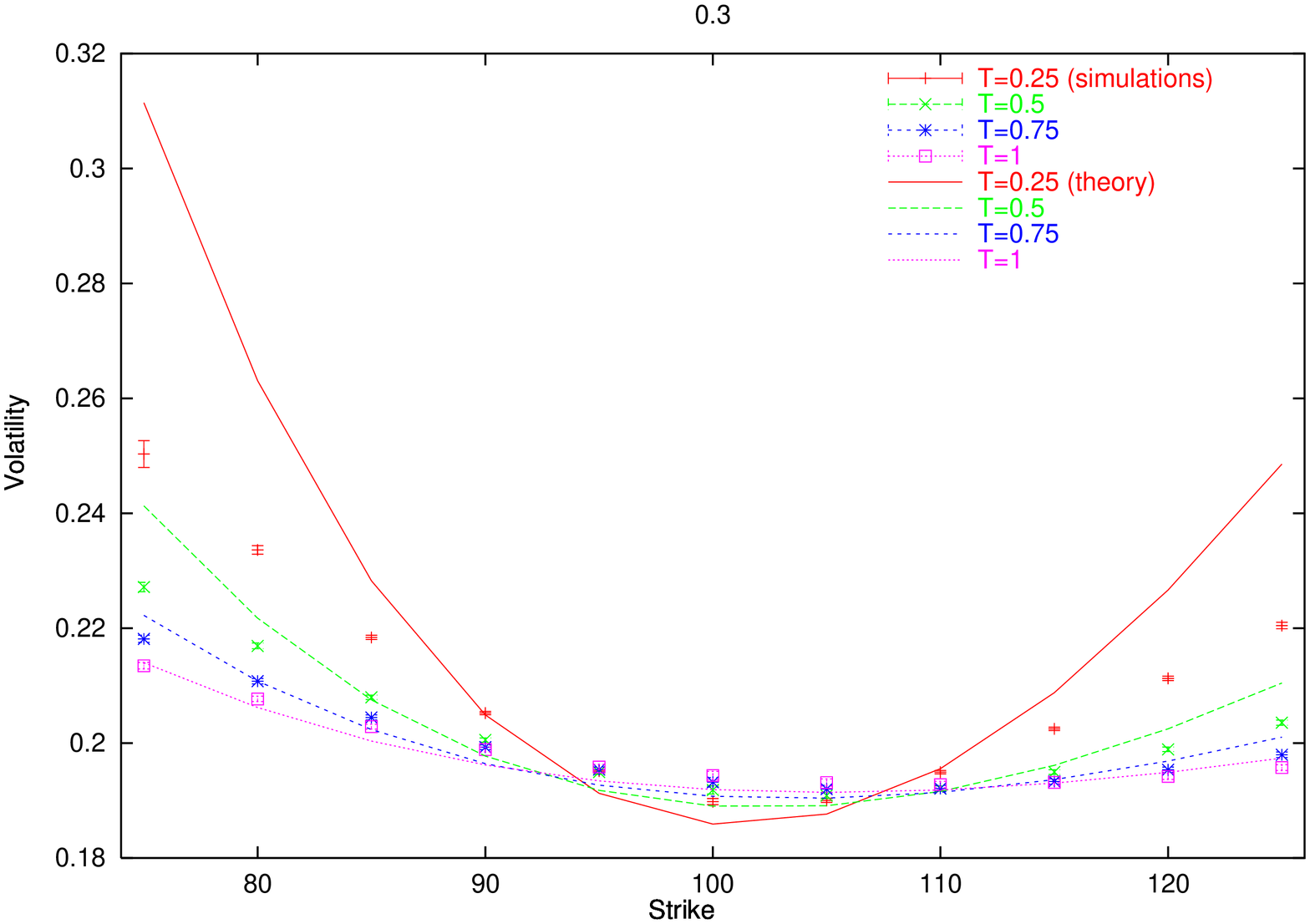} \includegraphics[angle=270,scale=0.26]{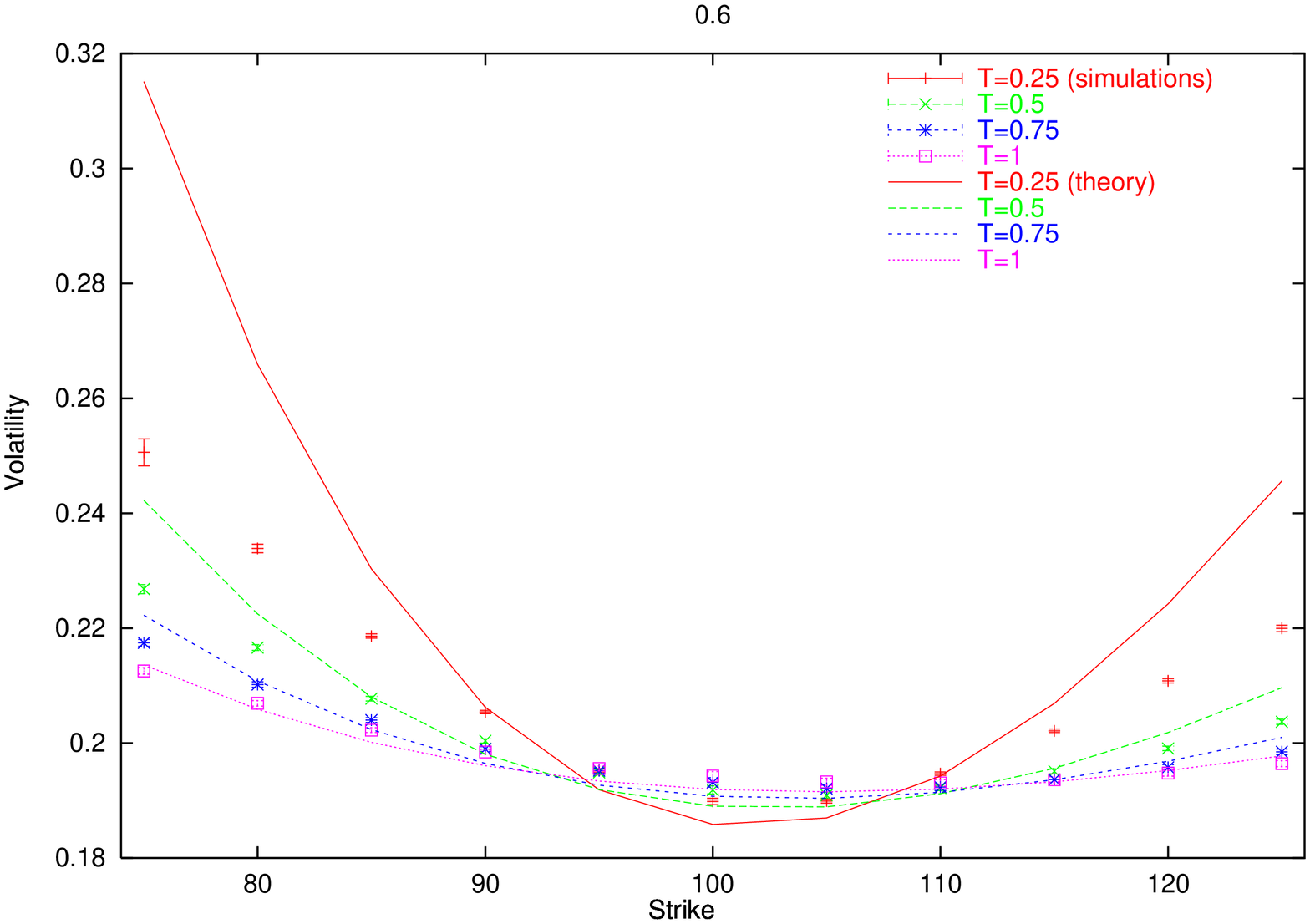}
    \caption{\label{impvol} Implied volatility $v_t$ as a function of the strike $K$,
for different maturities (t=0.25, 0.5, 0.75 and 1 year),
for $\alpha=0.3$ (left) and $\alpha=0.6$ (right). We have chosen $r=4\%$, $\sigma_1=20\%$, $\lambda=0.175$ and $T \simeq 2$ years (these values are chosen according to \cite{BMD}); the discretization step is here $\Delta t= 2^{-8}$. In both case we clearly see
the smile effect.}
  \end{center}
\end{figure}

\begin{figure}[!htbp]
  \begin{center}
    \includegraphics[scale=0.35]{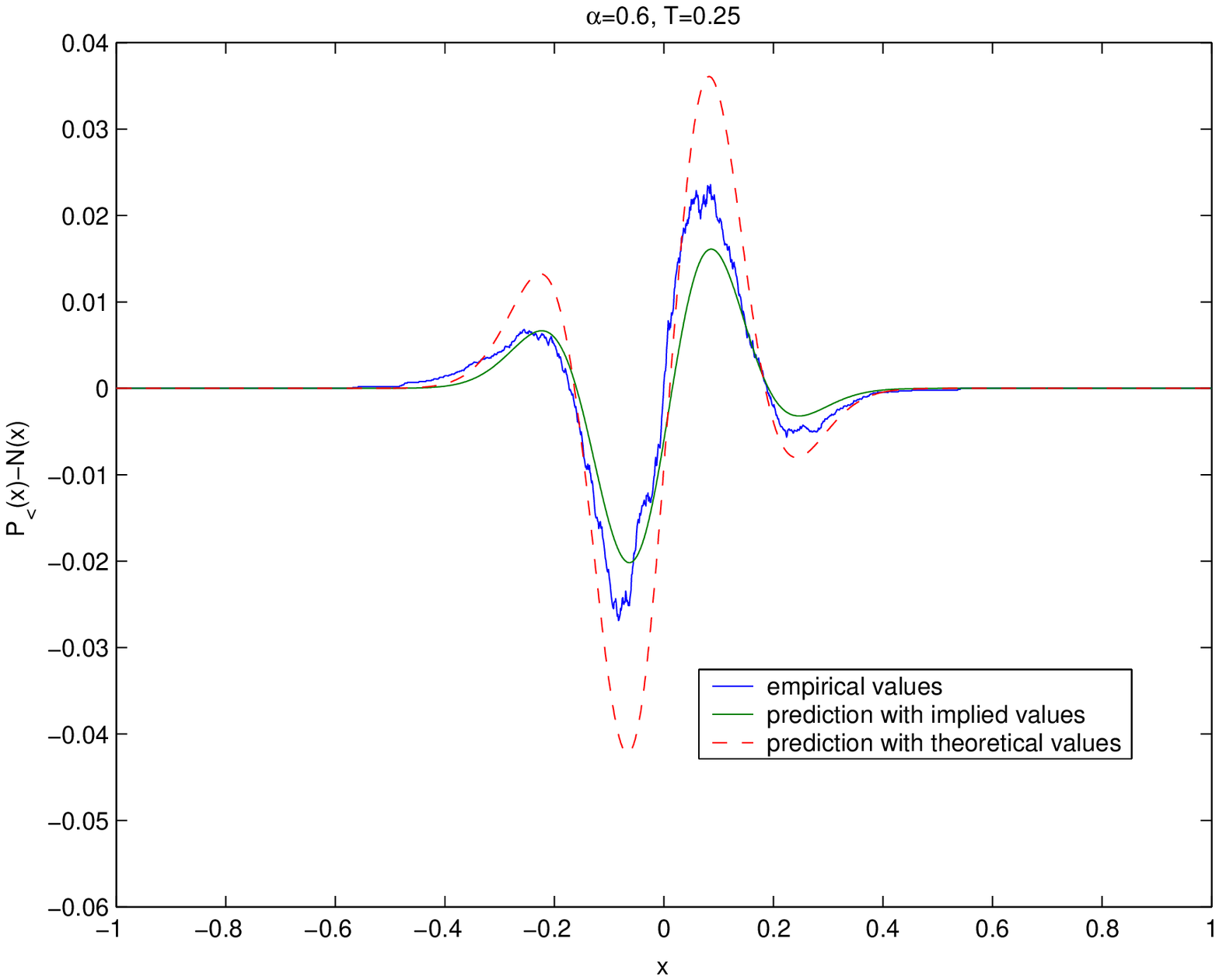} \includegraphics[scale=0.35]{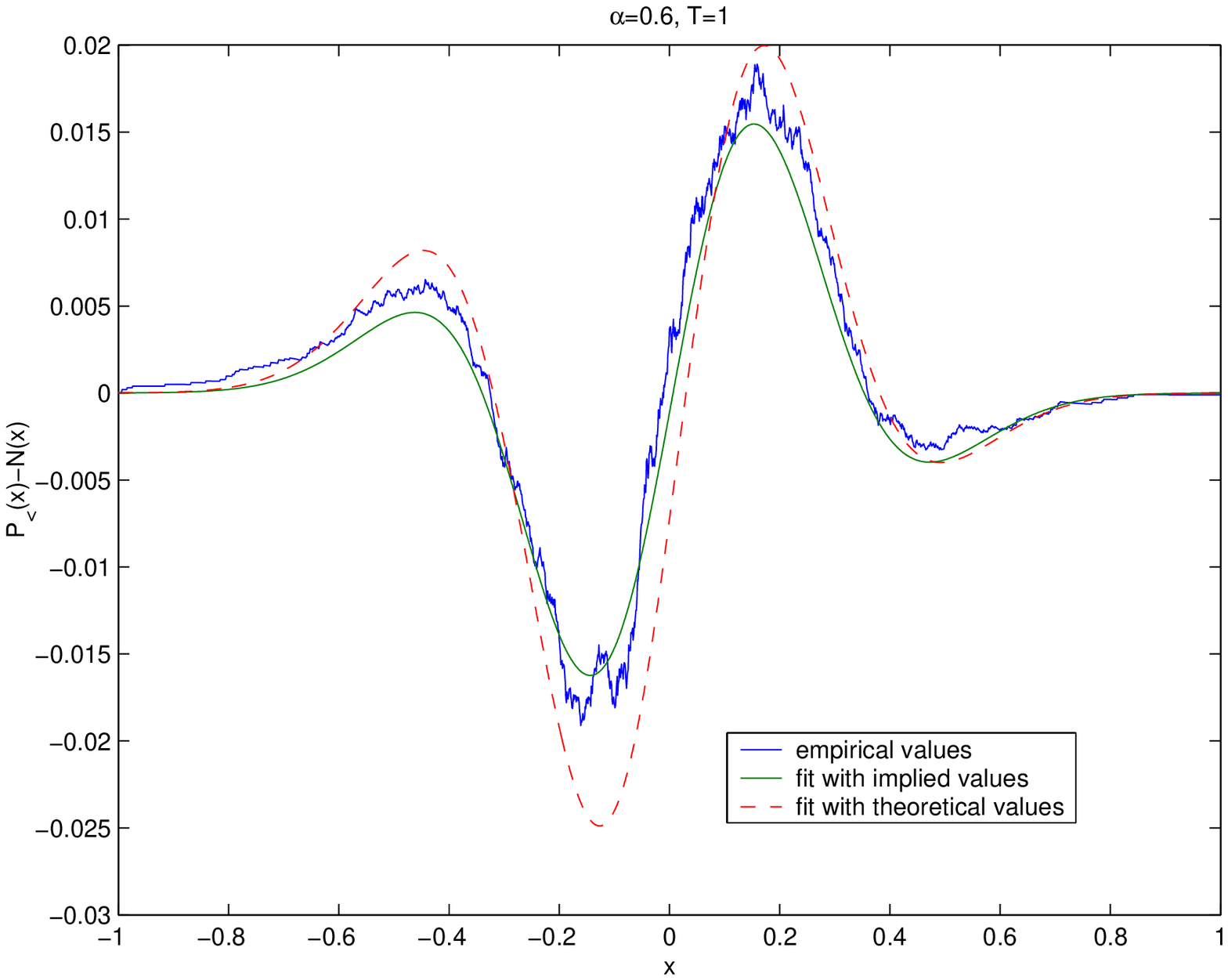}
    \caption{\label{edg} Difference between the distribution
        function of the {\sc smrw} ($\alpha=0.6$) and the normal distribution,
        for maturity 0.25 (left) and 1(right). We compare it
        with the cumulant expansion, evaluated with the implied values of the
        cumulants obtained by fit of the volatility smile and with
        the theoretical values of the cumulants(\ref{K3th},\ref{K4th}).
        The results are similar in the case $\alpha=0.3$. }
  \end{center}
\end{figure}

\begin{figure}[!htbp]
  \begin{center}
    \includegraphics[scale=0.35]{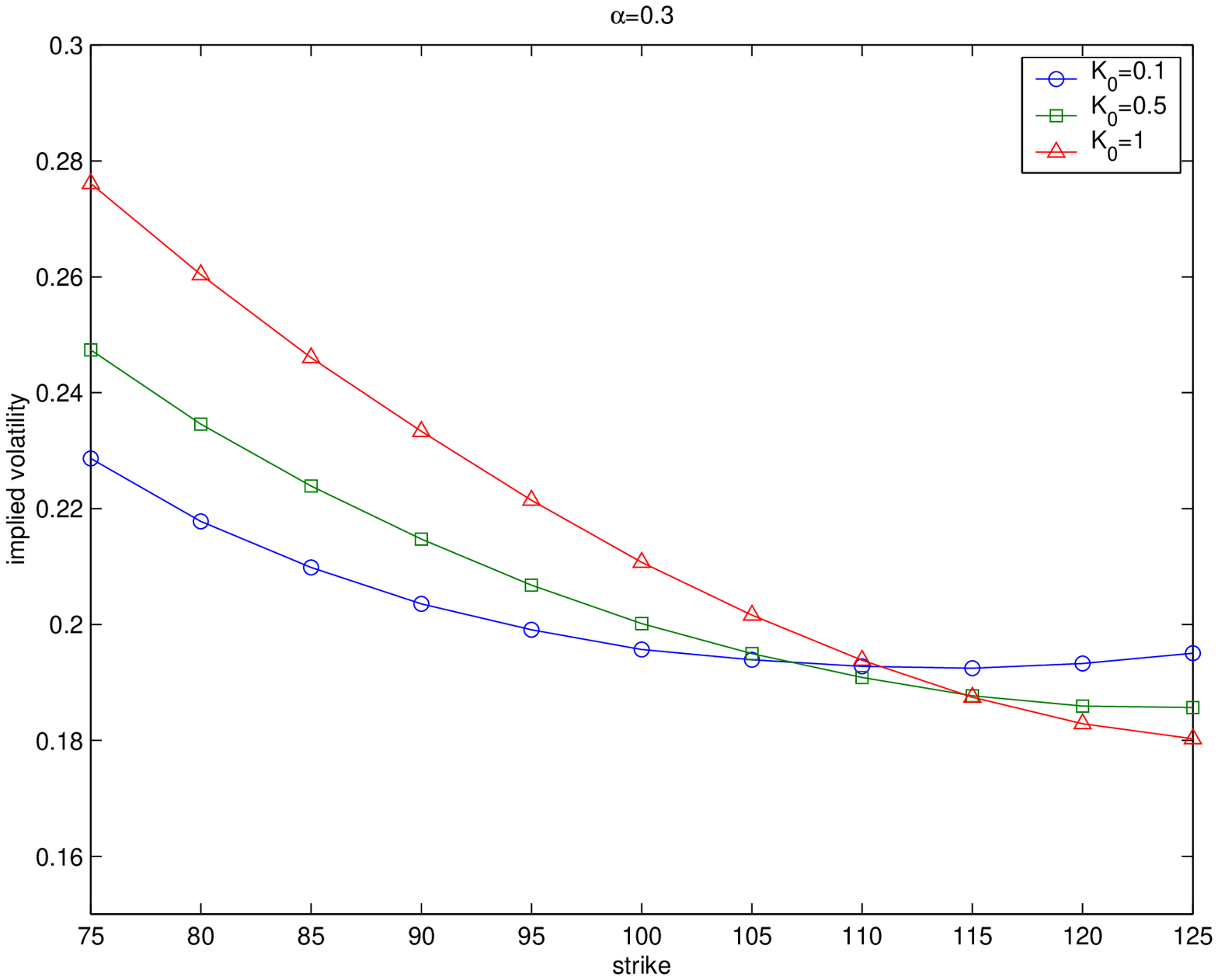} \includegraphics[scale=0.35]{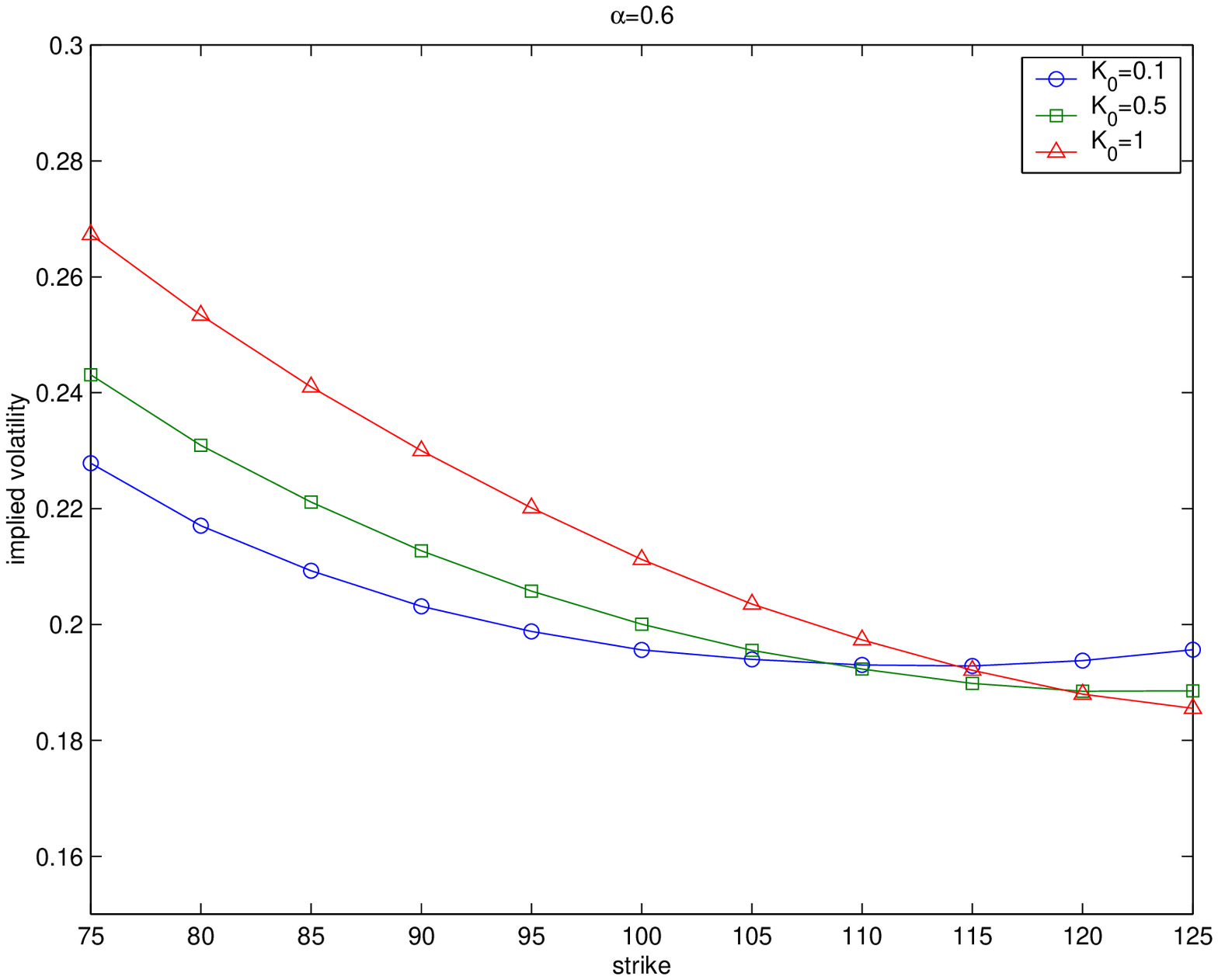}
    \caption{\label{smileK} Implied volatility $v_t$ as a function of the strike,
for different values of the parameter $K_0$, for $\alpha=0.3$ (left) and $\alpha=0.6$ (right).
We have chosen $r=4\%$, $\sigma_1=20\%$, $\lambda=0.175$. In both case we clearly see
the growing influence of $K_0$ on the asymmetric shape of the smile.}
  \end{center}
\end{figure}

It is a well known fact that Black-Scholes hypotheses are not satisfied in practice.
In particular, fat tail effects and volatility clustering are absent in the Black-Scholes
world. This has important
consequences on option pricing (particularly for \textit{exotic} options) and hedging.
One particularly clear symptom is the so called
\textit{volatility smile}: implied volatility from
option prices is not constant but varies across both strike and maturity, defining
a volatility surface.
Although this effect is widely known and studied by both academics and practitioners,
a detailed understanding of the implied volatility surface is
still one of the major challenge of modern finance.
\\Several authors \cite{jarrow,corrado,PCB,Backus} have given simple approximate formulas
for option
prices (or for the resulting volatility smile) when the asset is modelled by an arbitrary
stochastic process, which capture the basic mechanisms that lead to non trivial smiles.
The idea is to consider that the distribution of the price increments is
weakly non Gaussian. Using a truncated cumulant expansion \cite{Feller,jarrow}, one
finds corrections to the constant volatility case in terms of the skewness and kurtosis of the
underlying process (see also \cite{Fouque} for equivalent results but in a different language).
For example, when the log-price difference between times $0$ and $t$ has a
variance $\sigma_t$, skewness $\kappa_{3,t}$ and kurtosis $\kappa_{4,t}$,
the volatility smile $v_t$ is approximately
given by \cite{Backus}\footnote{The formula given in \cite{Backus} is in fact
slightly different in that it neglects terms in $\sigma_t$ compared to the moneyness $d_t$,
which we have kept below.}:
\begin{eqnarray}
v_t&=&\sigma_0\left[1+\frac{\kappa_{3,t}}{3!}(2\sigma_t-d_t)-
\frac{\kappa_{4,t}}{4!}(1-d^2_t+3d_t\sigma_t-3\sigma^2_t)\right] \label{smile}\\ \nonumber \\
d_t &=&\frac{\log(S_0/K)+rt+\sigma^2_t/2}{\sigma_t}, \nonumber
\end{eqnarray}
where $S_0$ is the current price of the underlying, $\sigma_0$ is (true) volatility,
$K$ the strike and $r$ the interest rate.
In the limit where $\sigma_t, rt \ll 1$ and $S_0-K \ll S_0$, the above formula boils
down to the
formula obtained in \cite{PCB} in the context of an additive model:
\begin{eqnarray}
v_t&=&\sigma_0\left[1-\frac{\kappa_{3,t}}{3!}d_t-
\frac{\kappa_{4,t}}{4!}(1-d^2_t)\right] \label{smile2}\\ \nonumber \\
d_t &=&\frac{S_0-K}{S_0\sigma_t}, \nonumber
\end{eqnarray}
Since our process displays both anomalous skewness and kurtosis,
we expect to observe this smile effect. The above formula is interesting
as it gives an explicit expression of the option smiles for the {\sc smrw} in terms of its
known skewness and kurtosis. It is therefore important to compare the result (\ref{smile})
with an exact smile obtained by Monte-Carlo simulations (Fig.\ref{impvol}), in order to
test the accuracy of the above formula. We model the returns of the underlying by a geometric
{\sc smrw}:
\begin{equation}
\frac{\delta S_t}{S_t}=\mu \delta t + \sigma_0 \delta X_t  \label{dynamique}
\end{equation}
where $X_t$ is a {\sc smrw}. We then evaluate the price of an (european)
option as an unconditional average of the final payoff $(S_T-K)_+=\max(S_T-K,0)$
over many realizations of our process with $\mu=r$. This assumes that
\begin{itemize}
\item the option is hedged using the Black-Scholes $\Delta$ \cite{Book,PBS},
which is a good approximation for the minimum variance hedging strategies. In this case,
the `risk neutral' process is indeed given by Eq.~(\ref{dynamique}).
Other hedging schemes are in fact possible, and would lead
to small corrections to the option price, but this is beyond the aim of the present analysis,
which is to test Eq.~(\ref{smile}) in the framework within which it is established.
\item one does not attempt to measure the instantaneous volatility and to condition the
paths to start from this measured value. In other words, we study the unconditional, time
independent volatility surface. Correspondingly, both the price and the
hedge are independent of the {\it current} level of volatility.
\end{itemize}
In Fig.\ref{impvol} we have drawn the volatility smiles given by the Monte-Carlo
simulation and the theoretical smile given by formula (\ref{smile}),
with the theoretical skewness $\kappa_{3,t}^{the}$ and kurtosis
$\kappa_{4,t}^{the}$ obtained from our above analysis:
\begin{eqnarray}
\kappa_{3,t}^{the}&=& -\frac{6K_0\sigma T^{\frac{3\lambda^2}{2}}
t^{\frac{1}{2}-\nu_2}}{(2-\nu_2)(1-\nu_2)}\Delta t^\mu \label{K3th}\\
\nonumber \\
\kappa_{4,t}^{the}&=& \frac{6T^{4\lambda^2}t^{-4\lambda^2}}
{(2-4\lambda^2)(1-4\lambda^2)}-3 \label{K4th}.
\end{eqnarray}
As already mentioned in section \ref{SEMRW}, the skewness grows with time when
$\alpha<\frac{1}{2}$.\footnote{At least in the regime $t \ll T$.
For $t \gg T$, the multifractal features disappear and both
the skewness and kurtosis vanish, as the process converges to a Gaussian.}
This can be of great interest for financial applications since it is often observed
that the skewness of the option smile can persist for large maturities.
We clearly see in (Fig.\ref{impvol}) that the non Gaussian nature of our process
induces an asymmetric smile and, visually, the agreement between simulations and
theory is satisfying for maturities $0.5,\,0.75,\,1$ and strikes between $80$ and $120$.
The case of the shortest maturity is less convincing, as expected from a cumulant
expansion, that is in principle only valid for long enough maturities.
In order to be more precise we then have performed a
parametric fit of these curves using  Eq.~(\ref{smile}) to obtain implied values of
the skewness ($\kappa_3^{imp}$) and the kurtosis ($\kappa_4^{imp}$).
We have chosen to fit the curves over the whole interval ($[75:125]$) of strikes.
It is not obvious that this is the best choice because the implied volatility for
deep in-the-money options (small strike) is known to be very sensitive to errors
(because of a small Vega).
We have also computed the empirical skewness and kurtosis from the simulations.
We report these results in Tab.\ref{impcum3} and Tab.\ref{impcum6}.
\begin{table}[!h]
\begin{center}
\begin{tabular}{|c|c|c|c|c|} \cline{2-5}
\multicolumn{1}{c|}{}&\multicolumn{4}{c|}{\textit{Maturity}} \\
\multicolumn{1}{c|}{}& 0.25 &0.5&0.75&1 \\ \hline
$\kappa_3^{imp}$ &-0.08 & -0.13 & -0.15& -0.16\\ \hline
$\kappa_3^{the}$ &-0.0975 & -0.107 & -0.114 & -0.118\\ \hline
$\kappa_3^{num}$ &-0.09 & -0.11 & -0.11 & -0.11\\ \hline \hline \hline
$\kappa_4^{imp}$ & 0.79 & 0.80 & 0.75 & 0.69 \\ \hline
$\kappa_4^{the}$ & 1.685 & 1.303 & 1.095 & 0.953 \\ \hline
$\kappa_4^{num}$ & 1.60 & 1.27 & 1.11 & 0.97 \\ \hline
\end{tabular}
\end{center}
\caption{\label{impcum3} Comparison between implied, theoretical and
numerically determined cumulants for different maturities, for $\alpha=0.3$.
The theoretical values are computed by formulas (\ref{K3th},\ref{K4th}).
The implied values are obtained by a fit of the Monte-Carlo values of volatility
using Eq.~(\ref{smile}), using all strikes.}
\end{table}
\begin{table}[!h]
\begin{center}
\begin{tabular}{|c|c|c|c|c|} \cline{2-5}
\multicolumn{1}{c|}{}&\multicolumn{4}{c|}{\textit{Maturity}} \\
\multicolumn{1}{c|}{}& 0.25 &0.5&0.75&1 \\ \hline
$\kappa_3^{imp}$ & -0.08 &-0.12&-0.14&-0.14\\ \hline
$\kappa_3^{the}$ & -0.136 &-0.122&-0.114&-0.109\\ \hline
$\kappa_3^{num}$ & -0.10 &-0.11&-0.10&-0.09\\ \hline \hline \hline
$\kappa_4^{imp}$ & 0.78 & 0.80 & 0.76 & 0.69 \\ \hline
$\kappa_4^{the}$ & 1.685 & 1.303 & 1.095 & 0.953  \\ \hline
$\kappa_4^{num}$ & 1.60 & 1.27 & 1.10 & 0.96  \\ \hline
\end{tabular}
\end{center}
\caption{\label{impcum6} Same as in Tab.\ref{impcum3}, for $\alpha=0.6$}
\end{table}
The agreement between the theoretical and implied values is only fair,
even if the order of magnitude is correct. The implied kurtosis is smaller
than the theoretical one for small maturities with a better agreement for large
maturities, whereas the implied skewness
tends to depart from its theoretical value for large maturities. Note that
the skewness {\it increases} with maturity, at variance with most theoretical
models in which it decreases as $1/\sqrt{t}$. Note also that
for a fixed range of strikes, the range of relative moneyness decreases with
maturity and the skewness becomes more difficult to ascertain precisely.
In order to track the origin of the observed discrepancy\footnote{Note that the agreement between empirical and theoretical
volatility is much better (see Fig. \ref{impvol}).}, we have directly
tested the cumulant expansion on the cumulative distribution of our process.
We show in Fig. \ref{edg} the difference between the numerical cumulative
distribution of the {\sc smrw} and the Gaussian distribution for two different time lags,
$t=0.25$ and $t=1$, for $\alpha=0.6$.
We see that the cumulant expansion truncated at second order is only qualitatively correct
in our case, since we observe systematic deviations between the empirical
values and the predictions. This is particularly true when one uses
the exact theoretical cumulants, and suggests that higher order cumulants
cannot be neglected. This is in fact expected, since higher
order cumulants, for example even ones, only decay with time as
$\kappa_{2p,t}\sim t^{-2p(p-1)\lambda^2}$ for the {\sc smrw},
instead of $t^{-p/2}$ for iid returns. For example, $\kappa_{6,t}/\kappa_{4,t}
\sim t^{-8\lambda^2} \sim t^{-0.25}$ in our case. (Actually, cumulants even
diverge beyond a certain order $p^*=1+1/4\lambda^2$, see \cite{BMD3}).
Of course, the fit of the cumulative distribution
with the implied values extracted from the option smile is better, since these
values try to correct the inadequacy of the truncated expansion.
The conclusion of this study is that a simple cumulant expansion for the
{\sc smrw} only gives a qualitative description of option smiles. In order to
get quantitative results, one should use Monte-Carlo paths. However, the scaling
properties of the model might allow one to find exact analytic solutions in some cases.
Finally, we want to show many typical smile shapes can be obtained with
our process by tuning its degree of asymmetry
(measured by the parameter $K_0$).
We have shown in (Fig.\ref{smileK}) three smiles obtained for different
values of $K_0$: $0.1, 0.5, 1$. In this example a more pronounced
skewness can be obtained. The term structure of the skewness and of the curvature
of the smile can be chosen by playing with the parameters $\alpha$ and
$\lambda^2$.
\section{Conclusion and prospects}
In this paper, we have generalized the construction of the multifractal
random walk ({\sc mrw}) of \cite{BMD} to take into account the asymmetric
character of the financial returns. Our first motivation was to introduce in this class
of models the observed correlation between past returns and future volatility,
in such a way that the scale invariance properties of the {\sc mrw} are preserved.
This correlation leads to a power-law skewness of the returns, and is in fact deeply
connected with the so-called ``{\sc harch}'' effect, or ``causal cascade''
observed in \cite{HARCH,AMS}. Such a process can be very useful in finance
since it captures most of the stylized facts of the returns: non Gaussian shape at short scales,
volatility clustering, long range dependence, and, with the present work, tunable skewness.
Possible applications in finance are numerous: pricing of options, risk management, etc.
As illustrated above,
the model considered here leads to very versatile shapes of volatility surfaces with
anomalous term structure. In particular, one
can account for smiles with skewness that is constant or even increases with maturity.
This would not
be possible without a slowly decaying return-volatility (leverage) correlation.
From a more academic point of view, we have not been able to construct a skewed multifractal
process which remains skewed in the continuous time limit.
Whether this is a fundamental limitation, or
that alternative constructions are possible, is an open question. It might be possible
to marry the
retarded volatility model of \cite{BMP} with the {\sc mrw} and obtain a skewed process in the
continuous time limit. Another promising path would be to study the precise connections
between the present model and the recent ``causal cascade'' construction of Calvet and Fisher,
that extends the work presented in \cite{Mandel} and might allow to introduce the skew in a
satisfactory way. We hope to answer these questions in future work \cite{work}.
\\
\textsc{Acknowledgments:}
We thank E. Bacry and J-F. Muzy for discussions at an early stage of this work, and
especially J-F. Muzy for having pointed out to us the link with the ``causal cascade''
effect. We also thank L. Calvet for comments about this work, A. Matacz and
M. Potters for many discussions on the leverage effect and on volatility smiles.
\bibliography{biblio_multifractal}

\begin{thebibliography}{10}

\bibitem{Lo}
A.~Lo.
\newblock Long term memory in stock market prices.
\newblock {\em Econometrica}, 59:1279, 1991.

\bibitem{Ding}
Z.~Ding and Granger C.
\newblock Modeling volatility persistence of speculative returns: a new
  approach.
\newblock {\em Journal of Econometrics}, 73:185--215, 1996.

\bibitem{Stanley}
Y.~Liu, P.~Cizeau, M.~Meyer, C.-K Peng, and H.-E Stanley.
\newblock Correlations in economic time series.
\newblock {\em Physica A}, 245:437, 1997.

\bibitem{Cont}
R.~Cont.
\newblock Empirical properties of asset returns: stylized facts and statistical
  issues.
\newblock {\em Quantitative Finance}, 1(2):223, 2001.

\bibitem{BMD}
J.-F. Muzy, J.~Delour, and E.~Bacry.
\newblock Modelling fluctuations of financial time series: from cascade process
  to stochastic volatility model.
\newblock {\em Eur. Phys. J. B}, 17:537, 2000.

\bibitem{Stanley2}
P.~Cizeau, Y.~Liu, M.~Meyer, C.-K Peng, and H.-E Stanley.
\newblock Volatility distribution in the {\sc s}\&p500 stock index.
\newblock {\em Physica A}, 245:441, 1997.

\bibitem{Mantegna}
S.~Miccich\`e, G.~Bonanno, F.~Lillo, and R.~N Mantegna.
\newblock Volatility in financial markets: stochastic models and empirical
  results.
\newblock cond-mat/0202527.

\bibitem{Mandel}
B.~Mandelbrot, A.~Fischer, and L.~Calvet.
\newblock A multifractal model of asset returns.
\newblock Cowles Foundation Discussion Paper \#1164, 1997.

\bibitem{BMDshort}
E.~Bacry, J.~Delour, and J.-F. Muzy.
\newblock Multifractal random walk.
\newblock {\em Physical Review E}, 64, 2001.

\bibitem{Book}
J.-P. Bouchaud and M.~Potters.
\newblock {\em Theory of financial risks}.
\newblock Cambridge University Press, 2000.

\bibitem{PCB}
M.~Potters, R.~Cont, and J.-P. Bouchaud.
\newblock Financial markets as adaptive systems.
\newblock {\em Europhysics letters}, 41(3), 1998.

\bibitem{Backus}
D.~Backus, S.~Foresi, K.~Lai, and L.~Wu.
\newblock Accounting for biases in black-scholes.
\newblock Working paper of NYU Stern School of Business, 1997.

\bibitem{Calvet1}
L.~Calvet and A.~Fisher.
\newblock Multifractality in asset returns: theory and evidence.
\newblock Forthcoming in \textit{Review of Economics and Statistics}.

\bibitem{Calvet2}
L.~Calvet and A.~Fisher.
\newblock Forecasting multifractal volatility.
\newblock {\em Journal of Econometrics}, 105:27--58, 2001.

\bibitem{leverage}
G.~Bekaert and G.~Wu.
\newblock Asymmetric volatility and risk in equity markets.
\newblock {\em The Review of Financial Studies}, 13:1--42, 2000.

\bibitem{BMP}
J.-P. Bouchaud, A.~Matacz, and M.~Potters.
\newblock Leverage effect in financial markets: the retarded volatility model.
\newblock {\em Physical Review Letters}, 87:228701, 2001.

\bibitem{perello}
J.~Perello and J.~Masoliver.
\newblock Stochastic volatility and leverage effect.
\newblock cond-mat/0202203.

\bibitem{Taqqu-Samoro}
M.S. Taqqu and G~Samorodnisky.
\newblock {\em Stable Non-Gaussian Random Processes}.
\newblock Chapman \& Hall, 1994.

\bibitem{intro_selfsimilar_processes}
P.~Embrechts and M.~Maejima.
\newblock An introduction to the theory of selfsimilar stochastic processes.
\newblock {\em International Journal of Modern Physics B}, 14:1399--1420, 2000.

\bibitem{BMD3}
E.~Bacry, J.~Delour, and J.-F. Muzy.
\newblock Modelling financial time series using multifractal random walks.
\newblock {\em Physica A}, 299:84, 2001.

\bibitem{Beran}
J.~Beran.
\newblock {\em Statistics for long-memory processes}.
\newblock Chapman \& Hall, 1994.

\bibitem{Numerical}
W.T. Vetterling, S.A. Teukolsky, and B.P. Press, W.H.and~Flannery.
\newblock {\em Numerical recipes in C: the art of scientific computing}.
\newblock Cambridge University Press, 1993.

\bibitem{turbulence}
J.-P. Laval, B.~Dubrulle, and S.~Nazarenko.
\newblock Non locality and intermittency in 3-d turbulence.
\newblock {\em Physics of fluids}, 13:1995, 2001.

\bibitem{AMS}
A.~Arneodo, J.-F. Muzy, and D.~Sornette.
\newblock "direct" causal cascade in the stock market.
\newblock {\em European Physical Journal B}, 2(2):277--282, 1998.

\bibitem{HARCH}
G.~Zumbach and P.~Lynch.
\newblock Heterogeneous volatility cascade in financial markets.
\newblock {\em Physica A}, 298(3-4):521--529, 2001.

\bibitem{Muzy-private}
J.-F. Muzy.
\newblock private communication.

\bibitem{jarrow}
R.~Jarrow and A.~Rudd.
\newblock Approximate option valuation for arbitrary stochastic processes.
\newblock {\em Journal of Financial Economics}, 10:347--369, 1982.

\bibitem{corrado}
C.J. Corrado and T.~Su.
\newblock Implied volatility skews and stock index skewness and kurtosis
  implied by s\&p 500 index option prices.
\newblock {\em The Journal of Derivatives}, pages 8--19, 1997.

\bibitem{Feller}
W.~Feller.
\newblock {\em An introduction to probability theory and its applications}.
\newblock Wiley, 1971.

\bibitem{Fouque}
J.-P. Fouque, G.~Papanicolaou, and K.R. Sircar.
\newblock {\em Derivatives in financial markets with stochastic volatility}.
\newblock Cambridge University Press, 2000.

\bibitem{PBS}
M.~Potters, J.-P. Bouchaud, and D.~Sestovic.
\newblock Hedged monte-carlo: low variance derivative pricing with objective
  probabilities.
\newblock {\em Physica A}, 289:517--525, 2001.

\bibitem{work}
B.~Pochart and J.-P. Bouchaud.
\newblock Work in progress.

\end{thebibliography}
\bibliographystyle{unsrt}
\appendix
\section*{Technical appendix}
The computation of the $2p+1^{th}$ moment of $X_t$ involves terms of the form
\begin{displaymath}
\sum_{0 \leq i_1, \dots, i_q <n}\langle \epsilon[i_1]^{\gamma_1}
\dots \epsilon[i_q]^{\gamma_q}e^{\gamma_1 \tilde{\omega}[i_1]+\dots+
\gamma_q \tilde{\omega}[i_q]}\rangle,
\end{displaymath}
with
\begin{eqnarray*}
\gamma_i &\in& \mathcal{N^*} \\
\sum_i\gamma_i&=&2p+1
\end{eqnarray*}
Writing more explicitly $\tilde{\omega}[i]$, we can expand these terms as:
\begin{displaymath}
\sum_{0 \leq i_1, \dots, i_q <n} \langle \epsilon[i_1]^{\gamma_1}\,e^{\mathcal{K}_1
\epsilon[i_1]}\rangle\dots \langle \epsilon[i_q]^{\gamma_q}\,e^{\mathcal{K}_q\epsilon[i_q]}
\rangle
\langle  e^{\gamma_1 \omega[i_1]+\dots+\gamma_q \omega[i_q]}\rangle
\prod_{-\infty<i<n, i \neq i_1,\dots, i_q}\langle  e^{\mathcal{K}_i\epsilon[i]}\rangle.
\end{displaymath}
The exponential terms $\epsilon[i]^{\gamma_i}e^{\mathcal{K_i}\epsilon[i]}$
follow from the definition of the $\tilde{\omega}_i$ and $\mathcal{K}_i$ are
complicated prefactors.
The above expression looks quite involved but as we search for the leading term,
we can use (\ref{conditions1})
in order to simplify it. In particular, to the first order in $K_0$,
$\langle  e^{\mathcal{K}_i\epsilon_i}\rangle\simeq 1$. We also have
\begin{eqnarray*}
\langle \epsilon[i]^{\gamma_i}\,e^{\mathcal{K_i}\epsilon[i]}\rangle&\simeq&
\langle \epsilon[i]^{\gamma_i}(1+\mathcal{K_i}\epsilon[i]+\dots\rangle \\
&\simeq & \left \{ \begin{array}{ll} \langle \epsilon[i]^{\gamma_i} & \textrm{if }
\gamma_i \textrm{ is even} \\
\mathcal{K_i}\epsilon[i]^{1+\gamma_i} & \textrm{if } \gamma_i \textrm{ is odd}
   \end{array}\right .
\end{eqnarray*}
Due to (\ref{conditions1}), we deduce that the lowest order term in $K_0$ is
found when a maximum number of $\gamma_i$ are even, that is all $\gamma_i$ except one
(because we look at the $2p+1^{th}$ moment) are even.
Finally, using basic properties of the Gaussian vector,\{$\omega_i$\}, we find that:
\begin{displaymath}
\langle  e^{\gamma_1 \omega[i_1]+\dots+\gamma_q \omega[i_q]}\rangle
=\prod_{1\leq k<\ell \leq q}\left(\frac{T}{(1+|i_k-i_\ell|)\Delta t}\right)^{\gamma_k \gamma_\ell
\lambda^2}
\left(\frac{T}{\Delta t}\right)^{\lambda^2(\frac{1}{2}\sum \gamma_i^2-\sum \gamma_i)}.
\end{displaymath}
We want to know which combination of the $\{\gamma_i\}$ leads to the dominating term.
One way to do this is to study the scaling with $\Delta t$ of the term corresponding
to a given combination of $\{\gamma_i\}$. Combining the previous computations, we
find that the exponent of $\Delta t$ is:
\begin{displaymath}
\Delta t^{1+(\alpha-\beta)+\lambda^2+p(1+2\lambda^2)-q-\frac{1}{2}\lambda^2
\sum_{i=1}^q \gamma_i^2}.
\end{displaymath}
In the limit $\Delta t \to 0$, only the smallest exponent will contribute.
The only parameters we can adjust are $q$ and $\sum_{i=1}^q \gamma_i^2$, with the constraint
$\sum_{i=1}^q \gamma_i = 2p+1$. The two extrema are:
\begin{itemize}
\item $q=2, \,\gamma_{i_1}=2p,\, \gamma_{i_2} =1$ which corresponds to an exponent:
\begin{displaymath}
\mu_2=(\alpha-\beta)+\frac{\lambda^2}{2}+(p-1)(1-2p\lambda^2).
\end{displaymath}
\item $q=p+1, \,\gamma_{i_1}=\dots=\gamma_{i_p}=2, \, \gamma_{i_{p+1}}=1$ which
corresponds to an exponent:
\begin{displaymath}
\mu_{p+1}=(\alpha-\beta)+\frac{\lambda^2}{2}.
\end{displaymath}
\end{itemize}
The difference $\mu_2-\mu_{p+1}=(p-1)(1-2p\lambda^2)$ is positive as soon as
$p<p^*=\frac{1}{2\lambda^2}$.
Therefore, we can conclude that the behaviour of the first $2p^*+1$ moments
is dominated by only one term,
corresponding to pairing all points except one. A similar discussion can be
made for even moments as well.
The above argument is clearer in the {\sc mrw} model \cite{BMD}, where the calculation is
easier and can be made without any further assumptions. In this model,
the odd moments are zero and the even moments have a scaling in $\Delta t$ given
by the exponent $p(1+2\lambda^2)-q-\frac{\lambda^2}{2}\sum_{i=1}^q\gamma_i^2$. We
immediately see that if $q=p \textrm{ and }\gamma_i=2$, this exponent is 0,
which is the condition for the existence of a non trivial continuous limit \cite{BMD}.
\end{document}